\documentclass{WileyMSP-template}
\newcommand{\fig}{Fig.\,}
\begin{document}

\raggedright

\title{3D printed micro-optics for quantum technology: Optimized coupling of single quantum dot emission into a single mode fiber}

\maketitle


\author{Marc Sartison$^{1, \dagger}$,}
\author{Ksenia Weber$^{2, \dagger}$,}
\author{Simon Thiele$^{3}$,}
\author{Lucas Bremer$^{4}$,}
\author{Sarah Fischbach$^{4}$,}
\author{Thomas Herzog$^{1}$,}
\author{Sascha Kolatschek$^{1}$,}
\author{Stephan Reitzenstein$^{4}$,}
\author{Alois Herkommer$^{3}$,}
\author{Peter Michler$^{1,*}$,}
\author{Simone Luca Portalupi$^{1,*}$,}
\author{and Harald Giessen$^{2,*}$}


\dedication{}

\begin{affiliations}
$^{1}$IHFG, Research Center SCoPE, and Integrated Quantum Science and Technology Center IQST, University of Stuttgart, Stuttgart, Germany\\
$^{2}$4$^{th}$ Physics Institute, Research Center SCoPE, and Integrated Quantum Science and Technology Center IQST, University of Stuttgart, Stuttgart, Germany\\
$^{3}$Institute of Applied Optics (ITO) and Research Center SCoPE, University of Stuttgart, Stuttgart, Germany \\
$^{4}$Institute of Solid State Physics, Technical University of Berlin, Berlin, Germany\\ 
\medskip
*Email address: h.giessen@pi4.uni-stuttgart.de, s.portalupi@ihfg.uni-stuttgart.de, p.michler@ihfg.uni-stuttgart.de \\
\medskip
$\dagger$ These authors contributed equally to this work. 

\end{affiliations}



\justifying
\begin{abstract}
\section{Abstract}
Future quantum technology relies crucially on building quantum networks with high fidelity. To achieve this challenging goal, it is of utmost importance to connect single quantum systems in a way such that their emitted single-photons overlap with the highest possible degree of coherence. This requires perfect mode overlap of the emitted light of different emitters, which necessitates the use of single mode fibers. Here we present an advanced manufacturing approach to accomplish this task: we combine 3D printed complex micro-optics such as hemispherical and Weierstrass solid immersion lenses as well as total internal reflection solid immersion lenses on top of single InAs quantum dots with 3D printed optics on single mode fibers and compare their key features. Interestingly, the use of hemispherical solid immersion lenses further increases the localization accuracy of the emitters to below $\SI{1}{nm}$ when acquiring micro-photoluminescence maps. The system can be joined together and permanently fixed. This integrated system can be cooled by dipping into liquid helium, by a Stirling cryocooler or by a closed-cycle helium cryostat without the necessity for optical windows, as all access is through the integrated single mode fiber. We identify the ideal optical designs and present experiments that prove excellent high-rate single-photon emission by high-contrast Hanbury Brown and Twiss experiments.\footnote{This paper is based in part on the PhD thesis by Marc Sartison \cite{PhD:Sartison2020}.}

\end{abstract}


\section{Introduction}

Single-photon emission from semiconductor quantum dots (QDs) has been shown to be a pure and efficient and non-classical light source with a high degree of indistinguishability, that could become an essential tool for future quantum communication \cite{Ates2009, He2013, somaschi2015, unsleber2015, gschrey2015, hanschke2018, Liu2019}.  In order make secure, wide-spread quantum communication networks \cite{duan2001, Gisin:2002aa, Gisin:2007aa} a reality, high fidelity single-photon sources are necessary. In the case of QDs, total-internal reflection (TIR) as a result of the high semiconductor-to-air refractive index contrast severely limits the single-photon extraction efficiency. Therefore, many attempts have been made to increase the extraction efficiency of quantum dots, using for example photonic crystal structures \cite{bensity1998}, cavity quantum electrodynamics \cite{gerard1998,  pelton2002, ates2012, somaschi2015, unsleber2015}, plasmonic surface effects \cite{biteen2006}, or solid immersion lenses (SILs) \cite{zwiller2002, ramsay2008, wildanger2012, sartison2017, bogucki2020}. Another crucial step in the development of practical quantum networks is the implementation of quantum repeater protocols, which enable long distance quantum communication via optical fiber channels. These protocols rely on the use of highly indistinguishable, entangled photons \cite{briegel1998, somaschi2015, thoma2016, Wang:2016aa} which requires the use of single-mode fibers \cite{bulgarini2014, Schlehan2018, Musiall2019}.  Thus, an efficient on-chip single-mode fiber coupled quantum light source is a key element to realize a QD based real-world quantum communication network. \\
In this paper, we first focus on enhancing the extraction efficiency of semiconductor QDs by optimizing micrometer-sized SILs designs. Two state-of-the-art technologies, namely low-temperature deterministic lithography \cite{sartison2017, dousse2008} and femtosecond 3D direct laser writing \cite{VonFreymann2010, Buckmann2012, fischer2013, Mueller2014, gissibl2016,Hohmann2015, gissibl2016two, Dietrich2017, jonuvsauskas2017, Dietrich2018} are used in combination to deterministically fabricate micro-lenses on pre-selected QDs. Due to the high flexibility of 3D direct laser writing, different SIL designs, namely hemispherical SILs (h-SILs), Weierstrass SILs (W-SILs) and total internal reflection SILs (TIR-SILs) can be produced and compared in regards to the single-photon extraction enhancement. The experimentally obtained values are compared to analytical calculations and the role of misalignment between SIL and QD as an error source is discussed in detail. We then highlight the implementation of an integrated single mode fiber coupled single-photon source based on 3D printed micro-optics \cite{bremer2020}. A 3D printed fiber chuck is used to precisely position an optical fiber onto a QD with a micro-lens printed on top. This fiber is equipped with another 3D printed lens designed to efficiently couple light from the TIR-SIL into the fiber core. Finally, we demonstrate that our compact on-chip solution is capable of producing an incoupling efficiency of $26\pm\SI{2}{\percent}$ into a single mode fiber.

\section{Results}
\subsection{Optical characterization of h-SILs}
\label{section:h-SILs}
\begin{figure}[!b]
	\centering
	\includegraphics[width=0.6\textwidth]{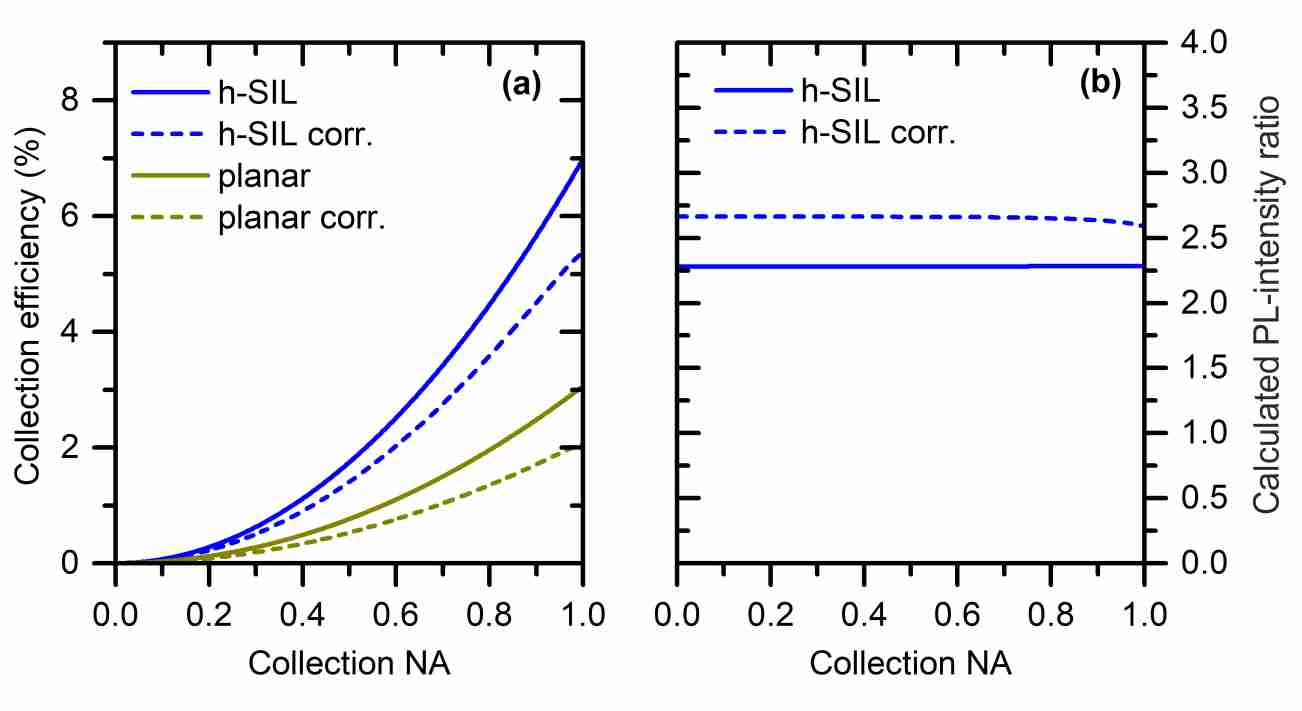}
	\caption{\textbf{(a)} Calculated collection efficiencies for planar extraction and extraction with an h-SIL  made from IP-Dip photoresist ($n=1.51$) placed centered on top of the emitter as a function of the detector collection NA. 
	\textbf{(b)} Calculated PL-intensity ratios for an h-SIL geometry over the used detector collection NA. The curves are obtained by division of the respective collection efficiency of the h-SIL by the planar one in (a).
	Solid curves represent ideal values, while dashed curves correct for one Fresnel reflection at the present interfaces (labeled as ``corr.''). 
	Here, refractive indices of $n_\mathrm{GaAs}=3.5$, $n_\mathrm{SIL}=1.51$, and $n_\mathrm{air}=1.0$ were used.}
	\label{fig:extractionefficiencies}
\end{figure}

In order to estimate the PL-intensity enhancement after the printing of an h-SIL, calculations were performed based on former works \cite{zwiller2002,barnes2002}.
Here, the QD was assumed to have a dipole emission characteristic which can be described in intensity as
\begin{align}
	I\left(\theta_1,\phi\right)\propto\frac{3}{8\pi}\left[1-\sin^2\left(\theta_1\right)\cos^2\left(\phi\right)\right]
	\label{eq:extraction_intensity}
\end{align}
using spherical coordinates.
The collection efficiency for a certain detector collection NA can be calculated by integration over the full  upper half solid angle:
\begin{align}
	\eta=\int_{0}^{\theta_{\mathrm{max}}} \mathrm{d}\theta_1 \int_{0}^{2\pi}	\mathrm{d}\phi ~	I\left(\theta_1,\phi\right)
\end{align}
$\theta_\mathrm{max}$ accounts for the total internal reflection (TIR) condition at the GaAs-to-air/lens interface and can be expressed as
\begin{align}
		\theta_\mathrm{max}^\text{no SIL} = \arcsin\left(\frac{n_\mathrm{air}}{n_\mathrm{GaAs}}\cdot NA\right)\\
	\theta_\mathrm{max}^\text{h-SIL} = \arcsin\left(\frac{n_\mathrm{SIL}}{n_\mathrm{GaAs}}\cdot NA\right)
\end{align}
for structures without SIL and with an h-SIL.
Numerical integration of equation \ref{eq:extraction_intensity} over the full upper half solid angle, while considering the conditions for $\theta_{\mathrm{max}}$, results in the solid collection efficiency curves displayed in \fig\ref{fig:extractionefficiencies}(a).
From these curves, a detector collection NA independent enhancement of $\tilde{\eta}_\mathrm{ideal}=2.28$ is derived by division of the upper solid curve by the lower solid one. 
This is shown in \fig\ref{fig:extractionefficiencies}(b) as the solid blue curve.
However, reflections at the interfaces are not taken into account, yet. This is performed by multiplying Eq.\,\ref{eq:extraction_intensity} with the according transmission Fresnel formula for each occurring refraction.
Since the polarization of the emitted light is mostly unpolarized, a transmission of
\begin{align}
	T_{total}=\frac{T_s+T_p}{2}
\end{align}
can be estimated for each interface. $T_s$ and $T_p$ represent the perpendicular and parallel polarized transmission with respect to the interface.
Thus, a reflection corrected intensity enhancement of $\tilde{\eta}_\mathrm{corr}\approx2.65$ (constant under NA variation) is calculated accounting for one reflection at each interface (see dashed blue curve in \fig\ref{fig:extractionefficiencies}(b)). All calculations are based on refractive indices of $n_\mathrm{GaAs}=3.5$, $n_\mathrm{SIL}=1.51$, and $n_\mathrm{air}=1.0$.\\

\begin{figure}[!h]
	\centering
	\includegraphics[width=0.6\textwidth]{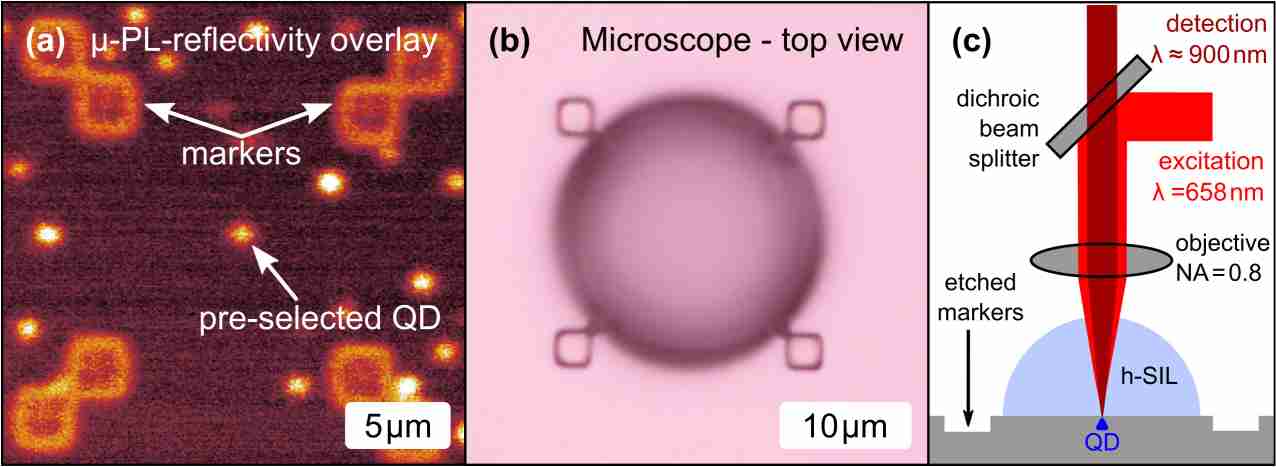}
	\caption{\textbf{(a)} High resolution overlay of a $\mu$-PL map and the corresponding reflectivity map (200 $\times$ 200 pixels) after the etching of the deterministically placed markers. The pre-selected QD appears to be well located in the center of the alignment structures. A longpass filter with a cut-off wavelength at \SI{865}{\nm} was used for suppressing the wetting layer signal. \textbf{(b)} Microscope top view of a deterministically 3D printed h-SIL with a diameter of \SI{20}{\um} aligned on the etched markers. \textbf{(c)} Simplified sketch of the $\mu$-PL setup. A laser diode emitting at \SI{658}{\nm} is used for QD excitation. The emitted light is transmitted via a dichroic beam splitter and coupled into a single-mode fiber which sends the photons to a spectrometer equipped with a CCD camera and a photon counting module. Figure reproduced from \cite{sartison2017}.}
	\label{fig:plmapmicroscope}
\end{figure}

For the sample processed here, markers are transferred into the substrate after resist development via ICP-RIE.
The accurate placement of the alignment markers with respect to the pre-selected QD can be seen via a $\mu$-PL scan after the marker fabrication in \fig\ref{fig:plmapmicroscope}(a). 
For data displayed in this figure, a high resolution $\mu$-PL map is merged with the simultaneously acquired reflectivity map provided by the excitation laser scan. 
\fig\ref{fig:plmapmicroscope}(b) shows a top view microscope picture of a deterministically printed lens with a diameter of \SI{20}{\um} which is placed with high accuracy centered on the QD. A quantitative study on this placement accuracy is discussed later in this work. 
For this sample, both deterministic lithography and characterization are carried out in a deterministic low-temperature lithography setup.
A schematic of the measurement principle is sketched in \fig\ref{fig:plmapmicroscope}(c). The QD is excited by a laser  at $\lambda_\mathrm{exc}=\SI{658}{\nm}$. 
Light emitted by the QD underneath the lens is collected by a $100\times$ low-temperature microscope objective ($\mathrm{NA}=0.8$) and passes the dichroic beam splitter (long pass \SI{675}{\nm}) while the reflection of red laser is reflected back into the excitation fiber.\\

 \begin{figure}[!h]
	\centering
	\includegraphics[]{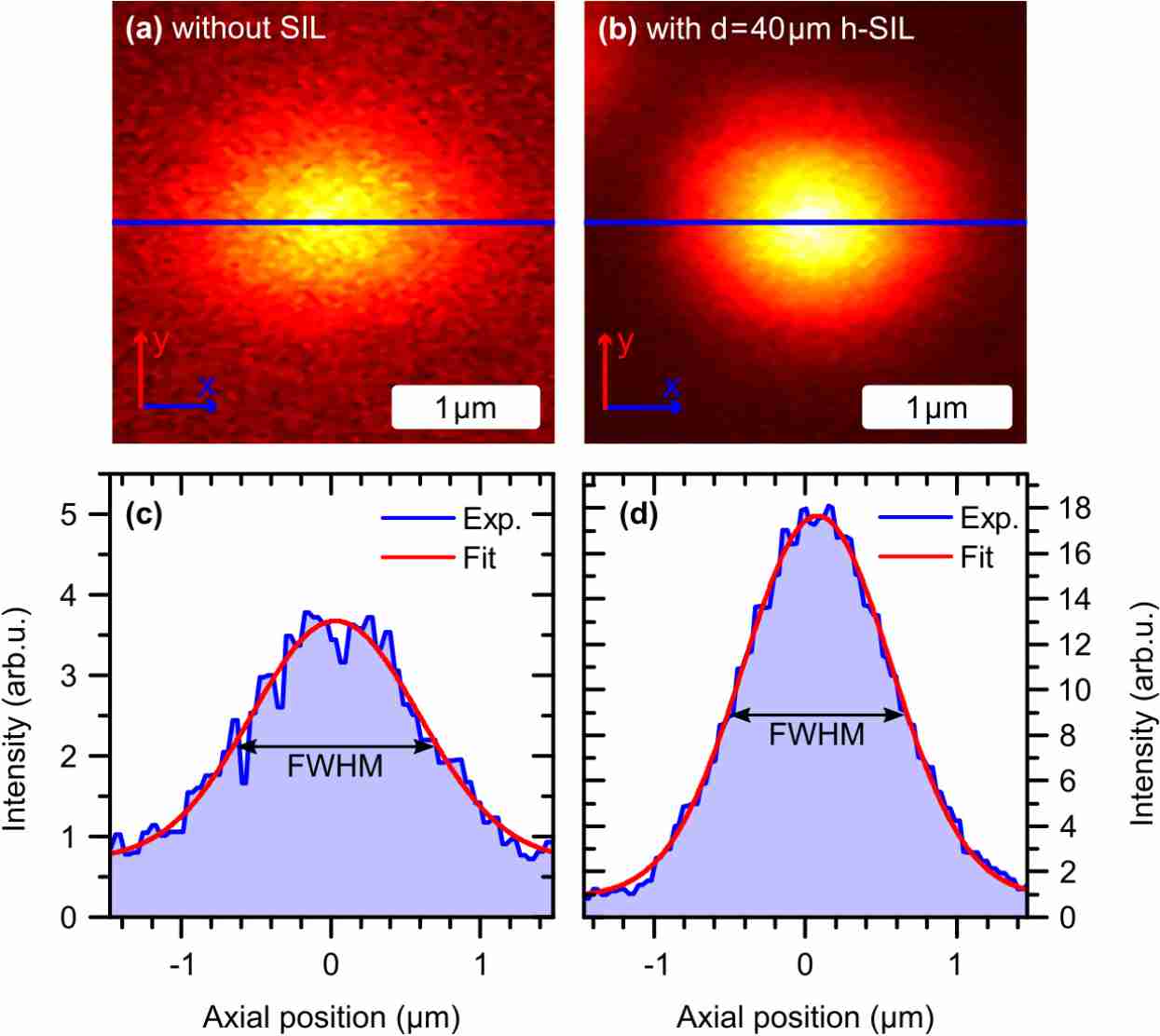}
	\caption{\textbf{(a)} and \textbf{(b)} High resolution \si{\micro}-PL maps of the same isolated QD without the h-SIL and after the fabrication respectively. A longpass filter with a cut-off wavelength of \SI{865}{\nm} was used for suppressing the wetting layer contribution. \textbf{(c)} and \textbf{(d)} Intensity profiles (blue) extracted of the maps from \textbf{(a)} and \textbf{(b)} (blue line) and corresponding Gaussian fits (red). A Gaussian fit is used for the signal being a convolution of the Gaussian laser beam and the QD spectrum. Figure analogous to \cite{sartison2017}.}
	\label{fig:mapsmitundohnelinse}
\end{figure}
In general, h-SILs are known for providing a better focusing of the laser beam \cite{gerardot2007,nickvamivakas2009, duocastella2015}. 
Due to the smaller spot size on the sample, a better signal-to-noise-ratio is observable when acquiring a \si{\micro}-PL intensity map. 
\fig\ref{fig:mapsmitundohnelinse}(a) and (b) show a high resolution \si{\micro}-PL map of the same single QD before and after the placement of an h-SIL. A longpass filter with a cut-off wavelength of \SI{865}{\nm} was used for suppressing the wetting layer contribution.
In order to obtain the QD position with respect to the system coordinates,
a 2D Gaussian surface fit was applied to the high resolution maps in \fig\ref{fig:mapsmitundohnelinse}(a) and (b). The error in the determination of the position of the maximum intensity is called localization accuracy.
By placing an h-SIL, an improvement in the horizontal scan direction from \SI{1.6}{\nm} to \SI{800}{\pico\meter} and in the vertical direction from around \SI{1.1}{\nm} to \SI{660}{\pico\meter} could be observed.
In summary, the h-SIL placement improves the localization accuracy by a factor of approximately 2.
It is worth mentioning that the QD localization only refers to the system coordinates and not to any processed markers or structure. 
Due to system specifications, an upper bound of the placement accuracy can be estimated to a conservative value of $\pm\SI{50}{\nm}$ which is negligible in comparison to the printed h-SIL size (diameter $\ge\SI{20}{\um}$).
Since the h-SIL reduces the spot size by a factor of $1/n_\mathrm{SIL}$ \cite{duocastella2015} (here $n_\mathrm{SIL}\approx1.51$) and provides a magnified image of objects underneath at the same time by a factor of $n_\mathrm{SIL}$, the FWHM should be mostly invariant whether an h-SIL is present or not.
This is verified by determination of the FWHM in the extracted cross-sections in \fig\ref{fig:mapsmitundohnelinse}(c) and (d). 
Respective values for the QD shown here are $\mathrm{FWHM}=1322\pm\SI{46}{\nm}$ and $\mathrm{FWHM_{h\text{-}SIL}}=1149\pm\SI{12}{\nm}$. \\

\begin{table}
	\centering
	{\begin{tabular}{|c|c|c|}
			\hline 
			\textbf{\begin{tabular}[c]{@{}c@{}}SIL\\diameter\end{tabular} (\si[detect-weight]{\um})} & \textbf{\begin{tabular}[c]{@{}c@{}}Measured\\displacement\end{tabular} (\si[detect-weight]{\nm})} & \textbf{\begin{tabular}[c]{@{}c@{}}Refraction-corrected\\displacement\end{tabular} $\Delta x_0$ (\si[detect-weight]{\nm})} \\ 
			\hline 
			20 & 1350 & 850 \\ 
			\hline 
			30 & 510 & 400 \\ 
			\hline 
			30 & 900 & 650 \\ 
			\hline 
			40 & 860 & 460 \\ 
			\hline 
			40 & 1210 & 680 \\ 
			\hline 
			50 & 330 & 220 \\ 
			\hline 
			50 & 1350 & 730 \\ 
			\hline 
			75 & 1270 & 740 \\ 
			\hline 
	\end{tabular} }
	
	\caption{Measured and refraction-corrected displacement values of printed h-SILs with different diameters with respect to the QD position. Table reproduced from \cite{sartison2017}. \fig\ref{fig:calcsbetter} illustrates in more detail how values for $\Delta x_0$ were obtained.}
	\label{tab:displacement_h-SIL}
\end{table}
\begin{figure}[!h]
	\centering
	\includegraphics[]{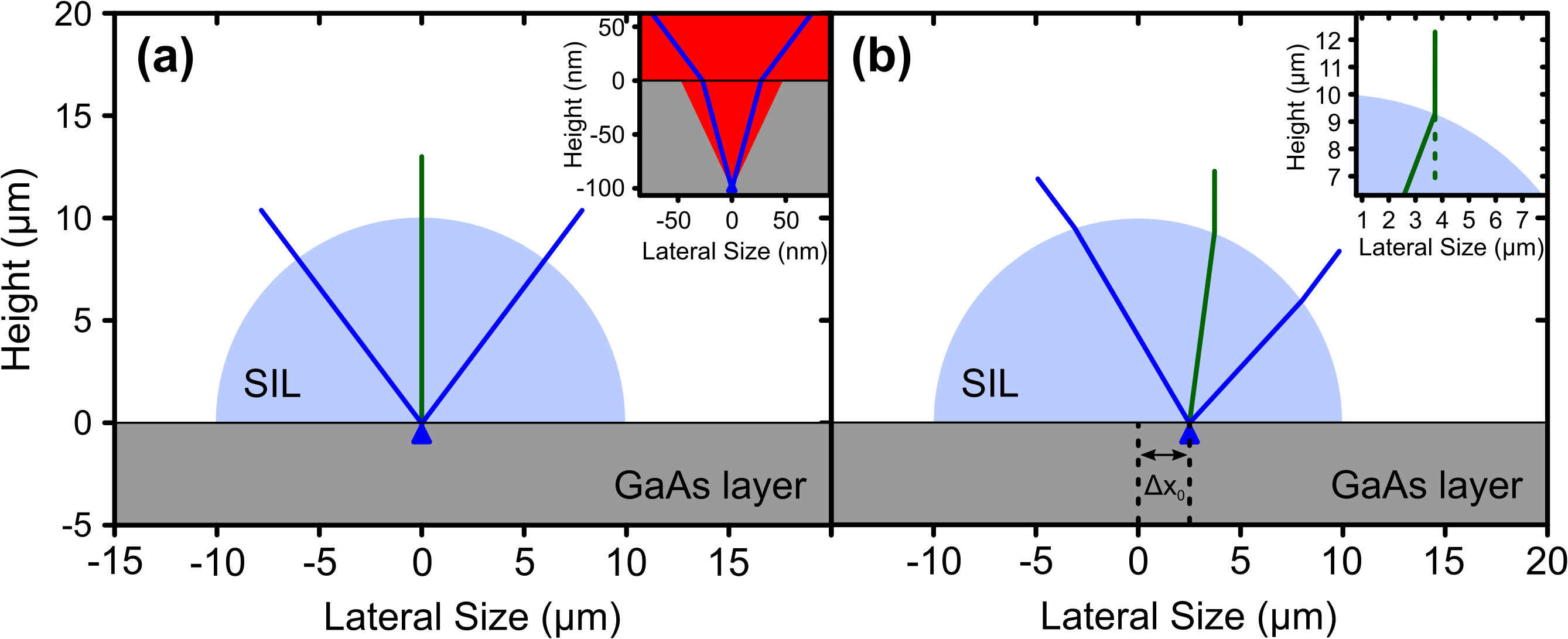}
	\caption{Illustration of the ray optics calculations to determine the displacement of the SIL. \textbf{(a)} The lens is perfectly centered with respect to the QD position. \textbf{(b)} The lens and the QD exhibit a lateral displacement of $\Delta x_0$. Optimizing the QD signal on the CCD camera yields a certain offset with respect to the lens center. This measurement allows for the calculation of the real QD displacement. Figure reproduced from \cite{sartison2017}.}
	\label{fig:calcsbetter}
\end{figure}
One question that remains is how precise the 3D printer can be aligned on the etched markers.
In order to tackle this question, the laser reflection pattern on top of the hemispheric SIL was imaged on the built-in camera module which appears circularly symmetric when aligned on the lens center. Furthermore, the distance to the maximum PL signal is recorded, which provides the measured displacement values in Tab.\,\ref{tab:displacement_h-SIL}.
However, these values do not represent the real lens displacement, since they do not account for refraction at the interfaces. 
To determine the refraction-corrected $\Delta x_0$ values in the last column in Tab.\,\ref{tab:displacement_h-SIL}, ray optics calculations were performed as illustrated in \fig\ref{fig:calcsbetter}.
\fig\ref{fig:calcsbetter}(a) shows ray optics calculations for a non displaced lens. The blue lines represent the fraction of light collected by the used detector collection NA. For \SI{0}{\nm} displacement, the central ray (here depicted in green) coincides with the ray which leaves the \ce{GaAs} and the h-SIL orthogonally. 
However, if the QD is displaced with respect to the lens center by $\Delta x_0$, the maximum PL signal is measured at a site which does not correspond to the real QD position, but rather to the position of the depicted green ray in \fig\ref{fig:calcsbetter}(b).
The obtained optimized coordinates now serve as a raw displacement input for further ray optics calculations which consider refraction at the interfaces.  Values obtained with this method are reported for different SIL diameters in Tab.\,\ref{tab:displacement_h-SIL}. 
These values demonstrate that a sub-micrometric SIL placement accuracy, in average around \SI{590}{\nm}, could be achieved with the fabrication method in this section. 
Since the laser used in the 3D printing phase is diffracted at the etched marker edges, it could be beneficial to switch to the deposition of metal markers to improve these values even more. In the following paragraphs, this placement accuracy is shown to be improved by the use of metal markers in combination with lens geometries more sensitive to spacial displacement: employing a similar approach of \fig \ref{fig:calcsbetter} the photoluminescence signal was already at its maximum when the beam was focused at the lens center, demonstrating a higher placement accuracy that the one achieved for h-SILs. \\

\begin{figure}[!h]
	\centering
	\includegraphics[]{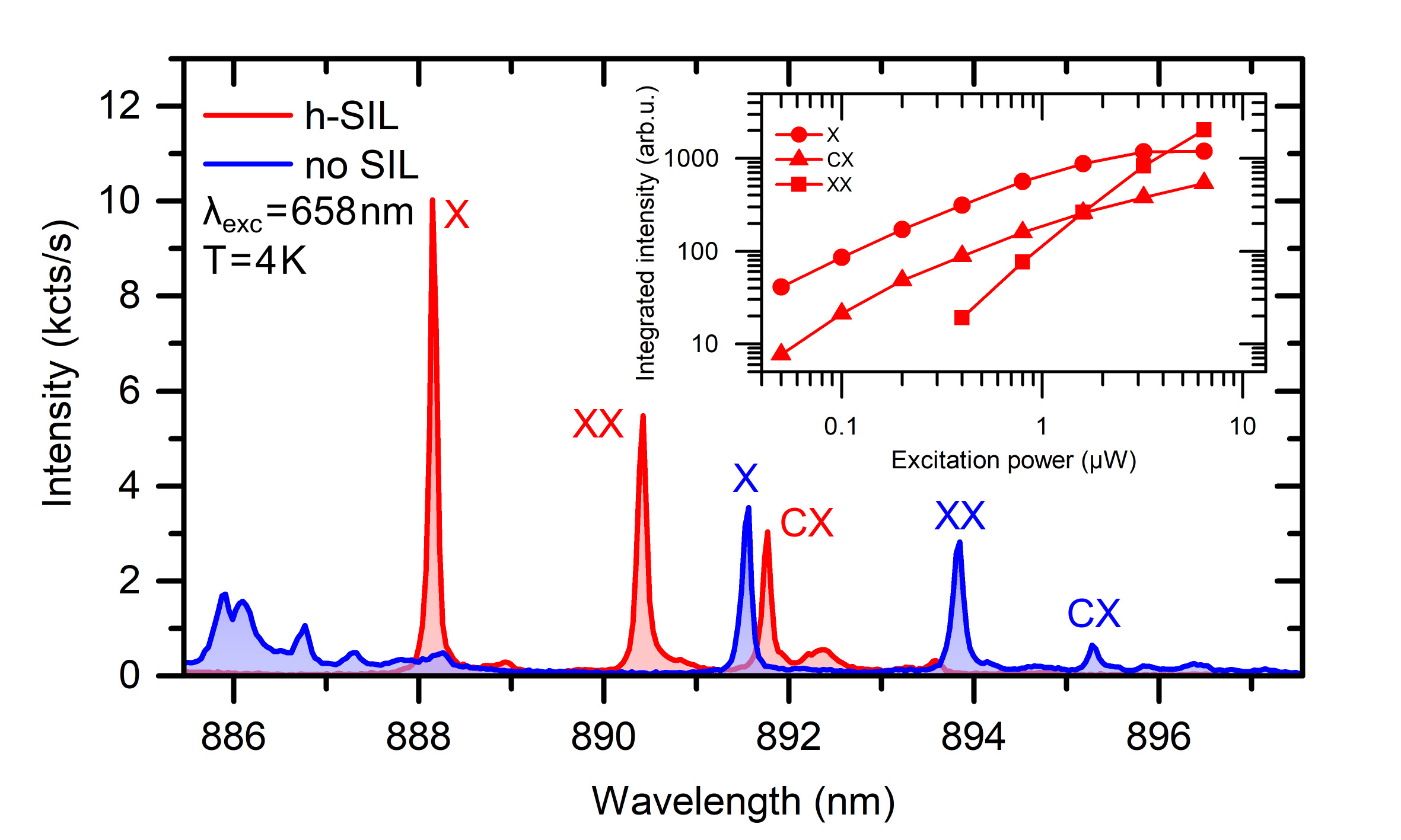}
	\caption{Exemplary spectrum of a QD without the lens before the fabrication (blue) and with the lens after the deterministic lithography and 3D printing (red). The inset shows a power-dependent measurements of the QD emission lines after the lens fabrication. Exciton (X) and charged exciton (CX) saturate at the same excitation power level, while the biexciton (XX) shows a superlinear behavior. A spectral blue shift of approximately \SI{3.5}{\nm} (\SI{5}{\milli\electronvolt}) after the SIL fabrication can be observed. Figure analogous to \cite{sartison2017}.}
	\label{fig:qd2spektrenundpowerserie}
\end{figure}
Changing the refractive index on top of the sample surface from $n=1$ to the SIL material $n\approx1.51$ leads to a change in the condition of total internal reflection. The critical angle is hereby enlarged from \SI{16.6}{\degree} to \SI{25.6}{\degree}, resulting in an increase of light extraction.
Indeed, the direct spectral comparison in \fig\ref{fig:qd2spektrenundpowerserie}
 before and after the h-SIL fabrication shows a broadband emission enhancement of all lines. This is observable in combination with an overall shift to higher emission energies. 
 \begin{figure}[!h]
 	\centering
 	\includegraphics[]{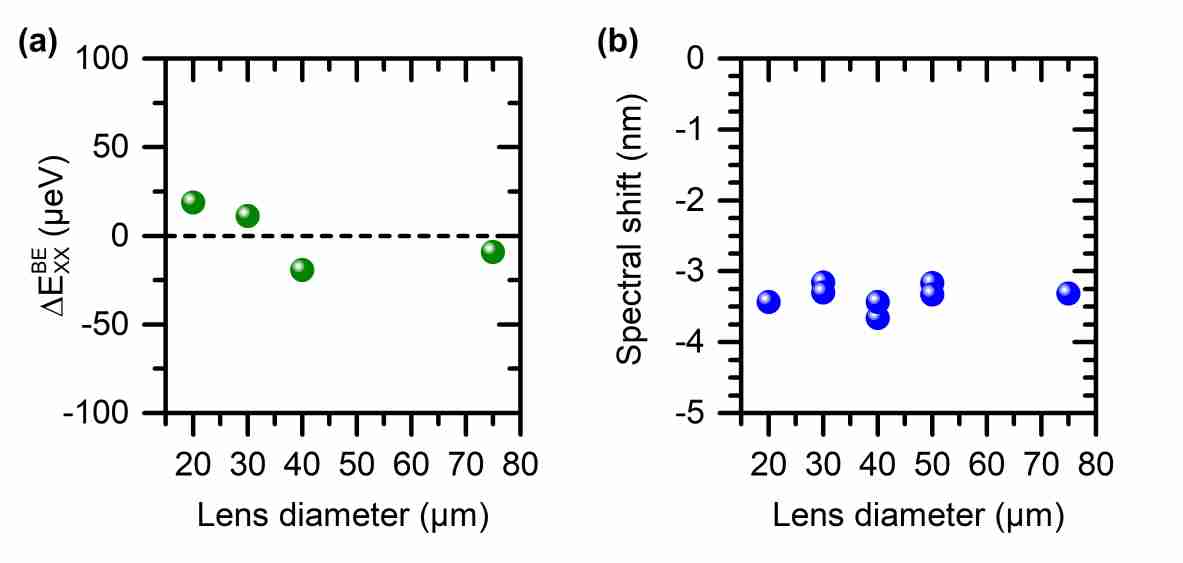}
 	\caption{\textbf{(a)} Change of the XX binding energy $\Delta E$ of the sample after the SIL fabrication and the cooling step to \SI{4}{\kelvin} for different lens diameters.
 		Not all lens diameters could be evaluated because of the absence of the XX for some QDs. \textbf{(b)} Spectral blue shift in dependence of the lens diameter when cooled down to \SI{4}{\kelvin}. A comparable blue shift for all fabricated h-SILs of around \SI{3.5}{\nm} is observed. For lens diameters of \SI{30}{\um}, \SI{40}{\um} and \SI{50}{\um}, two SILs were analyzed. Figure based on the same data as in \cite{sartison2017}.}
 	\label{fig:bindingshift}
 \end{figure}
Quasiparticle states are identified and labeled according to power dependent measurements which are depicted in the inset of \fig\ref{fig:qd2spektrenundpowerserie} for the sample after the SIL fabrication.
For this measurement, SILs are 3D printed at room temperature and then cooled down to \SI{4}{\kelvin} for the QD spectroscopy.
The overall blue shift can be attributed to local compressive strain being induced by the lens both as a result of the photoresist shrinking during the polymerization process and the polymerized resist having a bigger thermal expansion coefficient than the QD surrounding \ce{GaAs}. On average, all investigated QDs show a lens diameter independent spectral blue shift of approximately \SI{3.5}{\nm} (\SI{5}{\milli\electronvolt}), as shown in \fig\ref{fig:bindingshift}(b). 
This allows for the estimation of the applied stress resulting in a value of around \SI{180}{\mega\pascal} \cite{sartison2017, ding2010,trotta2015}.
When biaxial strain is applied on \ce{InAs} QDs, a variation of the XX binding energy is generally expected and also observed in the present case. \fig\ref{fig:bindingshift}(a) shows a variation of this value of around $\pm\SI{20}{\micro\electronvolt}$. \\

As already mentioned, the h-SIL leads to an increase of the light extraction due to the change in the TIR condition. However, the quantitative analysis is not obvious and requires further clarification, since the PL intensity ratio between different energy states of the QD might change after 3D printing of the SIL. 
This can be observed in \fig\ref{fig:qd2spektrenundpowerserie} for the X and the CX lines.
As already described, the h-SIL has a focusing effect on the excitation laser beam. The consequence is a different power density at the QD site. This can affect trapped charges in the QD vicinity, thus modifying the internal quantum efficiency \cite{sartison2017, hartmann2000}, 
which leads to a definition of the PL-intensity ratio of
\begin{align}
	\tilde{\eta} =\frac{I_\mathrm{X}^\mathrm{a}+\sum_i I_\mathrm{CX_i}^\mathrm{a}}{I_\mathrm{X}^\mathrm{b}+\sum_i I_\mathrm{CX_i}^\mathrm{b}}~.
	\label{eq:enhancementfactor}
\end{align}
$I_\mathrm{X}$ represents the integrated exciton intensity while $I_\mathrm{CX_i}$ stands for the integrated intensity of the i-th charged state. The superscripts a and b stand for intensities after and before the SIL fabrication.
Furthermore, the XX and higher transitions, if appearing, are not included in Eq.\,\ref{eq:enhancementfactor}. The enhancement factor stems only from transitions saturating at the same excitation power level which provide a reliable estimation of the PL-intensity ratio $\tilde{\eta}$. 
In order to exclude errors from the setup alignment between the measurements before and after the fabrication, a reference structure has always to be fabricated on the same chip. This reference must not have any surface modification on top of the QD.
By applying Eq.\,\ref{eq:enhancementfactor} on the reference QD in saturation, a correction factor $\alpha$ can be determined to account for small variations in the setup alignment.
The use of a reference emitter makes easy to compare results from different setups, even when using different collection NAs. Therefore, we can correct Eq.\,\ref{eq:enhancementfactor} to 
\begin{align}
	\tilde{\eta} =\frac{I_\mathrm{X}^\mathrm{a}+\sum_i I_\mathrm{CX_i}^\mathrm{a}}{I_\mathrm{X}^\mathrm{b}+\sum_i I_\mathrm{CX_i}^\mathrm{b}}\cdot \alpha~.
	\label{eq:enhancementfactor_real}
\end{align}
\begin{figure}[!h]
	\centering
	\includegraphics[]{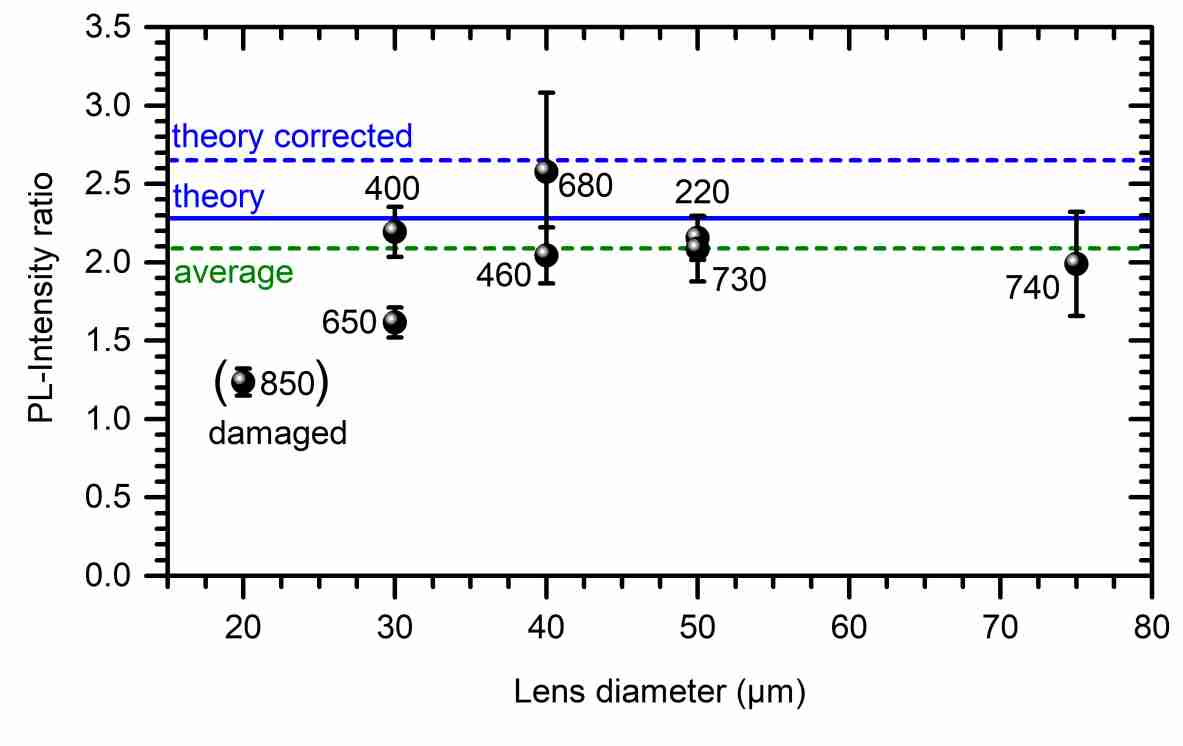}
	\caption{PL-intensity enhancement factor plotted over the SIL diameter. The numbers next to the data points represent the refraction-corrected displacement values from Tab.\,\ref{tab:displacement_h-SIL}. The green dashed line marks the average enhancement factor of 2.09. Corresponding theoretical calculations are shown in blue. The solid curve represents the expected PL-intensity ratios without considering reflections at the interfaces ($\tilde{\eta}=2.28$), while for the dashed line one reflection at each interface was considered ($\tilde{\eta}_\mathrm{corr}=2.65$). Figure based on the same data as shown in \cite{sartison2017}.}
	\label{fig:enhancementh-sils}
\end{figure}
By applying this data analysis to the spectra visible in \fig\ref{fig:qd2spektrenundpowerserie} it is possible to extract a value for this fabricated h-SIL (diameter \SI{30}{\um}) of $\tilde{\eta}=2.19\pm0.16$.
In this case the correction factor was found to be $\alpha=0.87$.
All lenses mentioned in Tab.\,\ref{tab:displacement_h-SIL} were investigated in the same manner.
The corresponding PL-intensity ratios are depicted in \fig\ref{fig:enhancementh-sils} and labeled with the according placement error.
An average PL-intensity ratio of $\tilde{\eta}=1.99\pm0.21$ could be determined. \\

\begin{figure}[!h]
	\centering
	\includegraphics[]{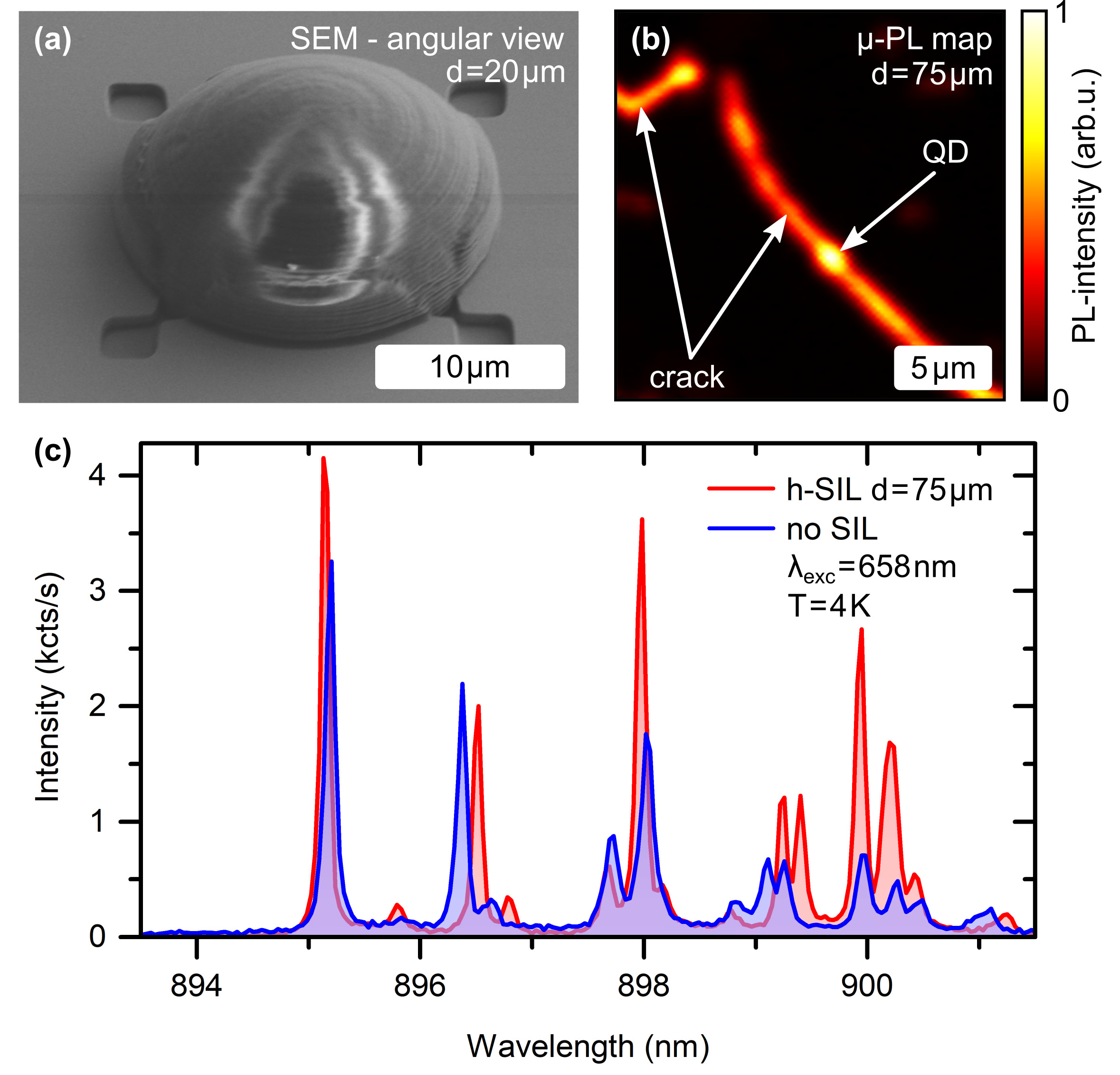}
	\caption{\textbf{(a)} SEM image of the excluded broken h-SIL in \fig\ref{fig:enhancementh-sils}. Parts of the lens are cracked and peeled off, most likely due to the cooling cycle to \SI{4}{\kelvin}. \textbf{(b)} $\mu$-PL intensity map of QDs underneath a cracked lens. The cracks are visible due to scattering of emitted QD light. Cracks most likely occur at the SIL-GaAs interface because of the lens shrinkage when cooled to cryogenic temperatures. A \SI{865}{\nm} longpass filter was used for suppressing the wetting layer signal. \textbf{(c)} Spectral comparison of a QD underneath a cracked lens with a diameter of \SI{75}{\um} at \SI{4}{\kelvin}. Spectral shift and PL-intensity enhancement are not observable here.}
	\label{fig:brokenlens}
\end{figure}

Further structural investigations pointed towards the smallest h-SIL being damaged, thus not delivering a perfectly hemispherical shape (see SEM image in \fig\ref{fig:brokenlens}(a)). 
When excluding the measurement point from the data evaluation in \fig\ref{fig:enhancementh-sils}, the average enhancement factor is found to be $\tilde{\eta}=2.09\pm0.23$ (indicated via the dashed green line). This matches  the theoretically calculated value of 2.28 (solid blue line) quite well, assuming perfect transmission at the interfaces.
The deviation from the calculated reflection corrected factor (dashed blue line) can be attributed to  uncontrolled diffraction due to surface roughness. 
These  surface variations are in the order of several tens of \si{\nm} and are thus not completely negligible. This can negatively influence the lens performance. 
It is also worth mentioning that the smallest lens was the only small one which got partially damaged. 
Large lenses exceeding diameters of \SI{75}{\um} and above are not suitable for low-temperature experiments, since they crack during the cooling because of the induced amount of strain at the lens-GaAs-interface. 
This leads to defects or a slight detachment, which results in the SIL not working properly anymore. 
An example \si{\micro}-PL scan of a QD underneath a cracked lens can be observed in \fig\ref{fig:brokenlens}(b).
Here, light emitted by QDs is scattered at the cracks which limits the SIL performance.
A further indication for the lens detachment is the absence of the expected spectral blue shift which is shown in a spectral comparison (\fig\ref{fig:brokenlens}(c)) for the same lens as investigated in \fig\ref{fig:brokenlens}(b). 
\subsection{Optical characterization of Weierstrass SILs}
\label{section:WSILs}
\begin{figure}[!h]
	\centering
	\includegraphics[]{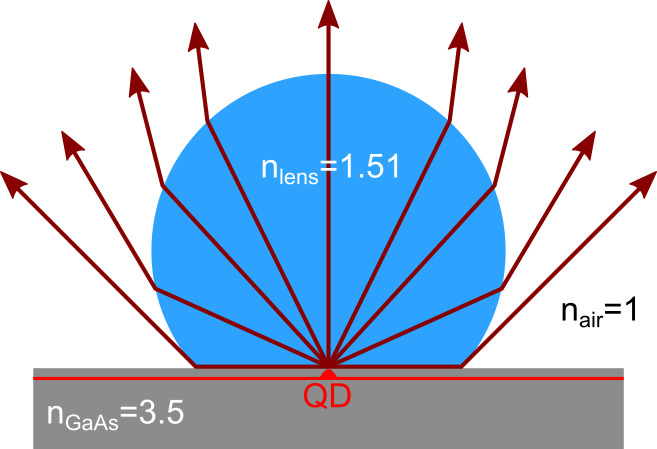}
	\caption{Schematic of a SIL in the Weierstrass geometry. All rays leaving the semiconductor are redirected into a output NA which is only dependent on the refractive index of the lens material.}
	\label{fig:weierstrasssilskizze}
\end{figure}

\begin{figure}[!h]
	\centering
	\includegraphics[]{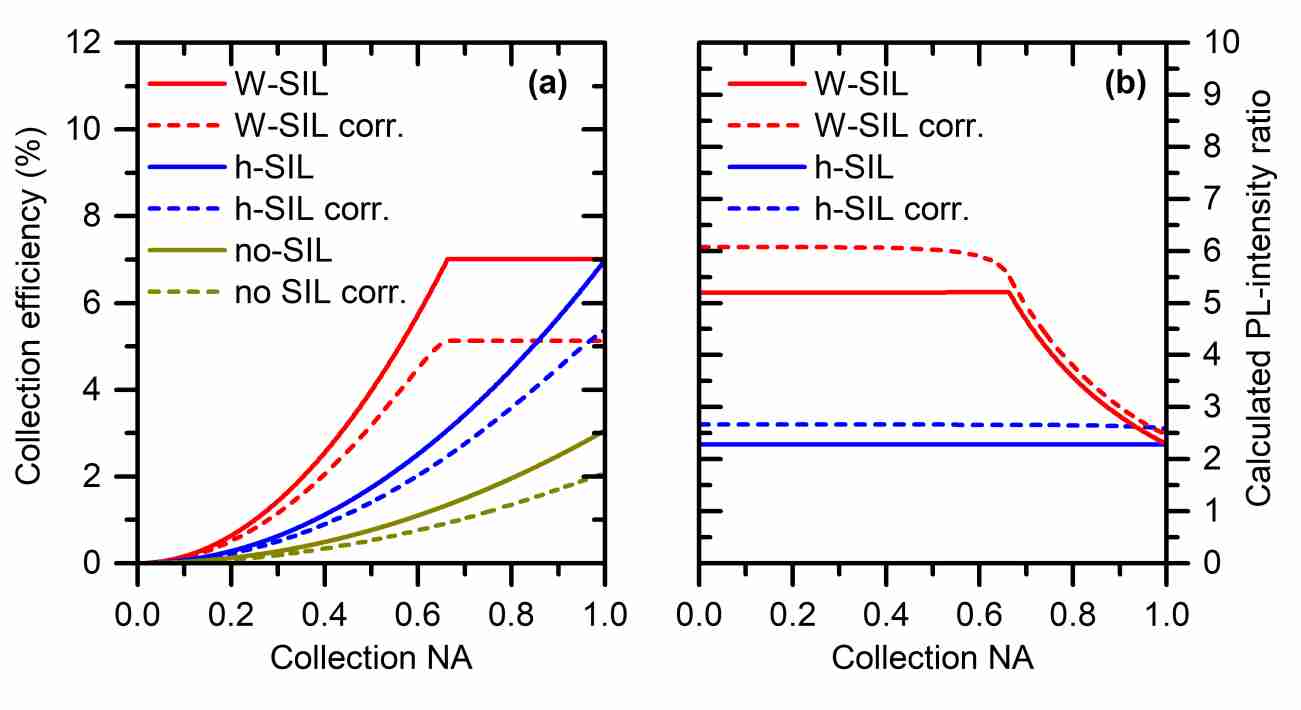}
	\caption{\textbf{(a)} Calculated collection efficiencies for planar extraction and extraction from a W-SIL or an h-SIL placed centered on top of the emitter as a function of the detector collection NA. 
		\textbf{(b)} Calculated PL-intensity ratios for a W-SIL and an h-SIL geometry versus the detector collection NA. The curves are obtained by division of the respective collection efficiency with the W-SIL by the planar one in (a).
		Solid curves represent ideal values, while dashed curves correct for one reflection at the present interfaces (labeled as ``corr.''). 
		Here, refractive indices of $n_\mathrm{GaAs}=3.5$, $n_\mathrm{SIL}=1.51$, and $n_\mathrm{air}=1.0$ were used.}
	\label{fig:extraction-efficiencies-w-sil}
\end{figure}

The investigated h-SILs exhibit good intensity enhancement. However, there are geometries which are much more promising. 
To boost the collection efficiency even further, in the second device generation the lens geometry has been changed to the so-called Weierstrass or hypersperical geometry. In this case, the lens has the shape of a truncated sphere with a height of $R\left(1+1/{n_\mathrm{SIL}}\right)$, with $n_\mathrm{SIL}$ representing the refractive index of the lens material. 
A schematic of the Weierstrass geometry used in this thesis with {$n_\mathrm{SIL}=1.51$} can be seen in \fig\ref{fig:weierstrasssilskizze}.  
Performing calculations in analogy to those previously discussed leads to the collection efficiencies displayed in \fig\ref{fig:extraction-efficiencies-w-sil}(a). 
A maximum integration angle for the Weierstrass SIL (W-SIL) \cite{zwiller2002} can be expressed as
\begin{align}
		\theta_\mathrm{max}^\text{W-SIL} = \arcsin\left(\frac{n^2_\mathrm{SIL}}{n_\mathrm{GaAs}}\cdot NA\right)~.
\end{align}
The curves of the W-SILs can be interpreted in the way that all the light which exits the semiconductor and enters the lens is folded into a certain output NA which is only dependent on the refractive index of the Weierstrass lens. 
For a refractive index of $n_\mathrm{SIL}=1.51$ this critical NA value results to be 0.66. This is verified when taking a closer look at the non-reflection corrected curves in \fig\ref{fig:extraction-efficiencies-w-sil}. 
Here, both solid curves of h- and W-SIL intersect at $\mathrm{NA}=1.0$, yielding the same overall collection efficiency.
The refraction-corrected curve ends up at a slightly lower value than for the h-SIL, which is expected. For an h-SIL, all angles of incident at the lens-to-air interface are in good approximation zero which leads to the highest possible transmission. As \fig\ref{fig:weierstrasssilskizze} illustrates, this is not the case for the W-SIL where strong refraction leads to significant Fresnel losses.
\begin{figure}[!h]
	\centering
	\includegraphics[]{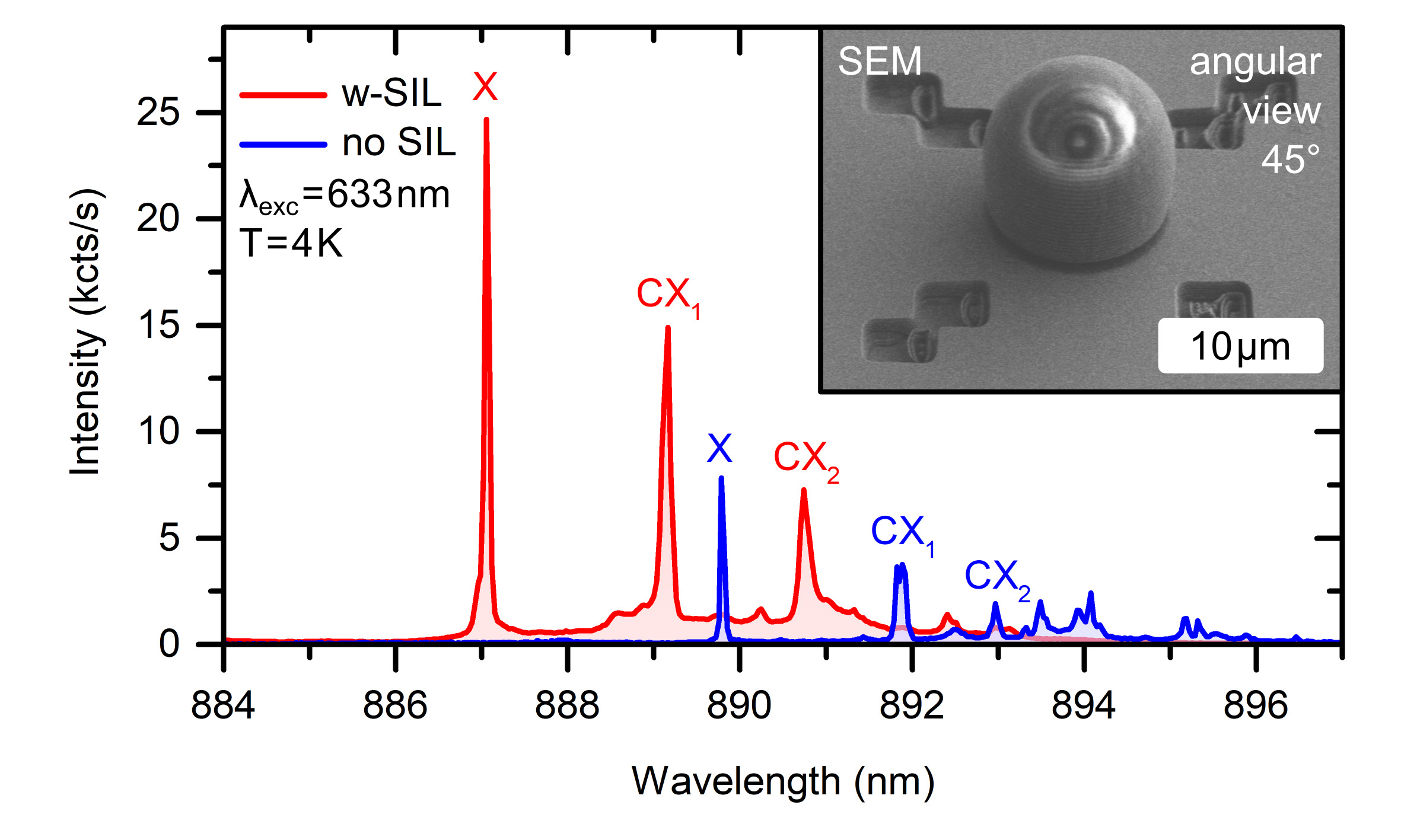}
	\caption{\si{\micro}-PL spectra of the same QD underneath a Weierstrass lens with a diameter of \SI{10}{\um} and without a lens. Emission characteristics were identified prior to the intensity enhancement evaluation via power-dependent measurements. The inset depicts a SEM angular view picture (\SI{45}{\degree} tilt) of the printed W-SIL.}
	\label{fig:weierstrasssilthesis}
\end{figure}
This points towards expected PL-intensity enhancement values of $\tilde{\eta}_\mathrm{ideal}=5.20$ and $\tilde{\eta}_\mathrm{corr}=6.02$ for detector collection NAs between 0 and 0.66 shown in \fig\ref{fig:extraction-efficiencies-w-sil}(b). For higher NAs, the PL-intensity ratio $\tilde{\eta}$ converges towards the ratio for h-SILs, as expected.\\

\begin{figure}[!h]
	\centering
	\includegraphics[]{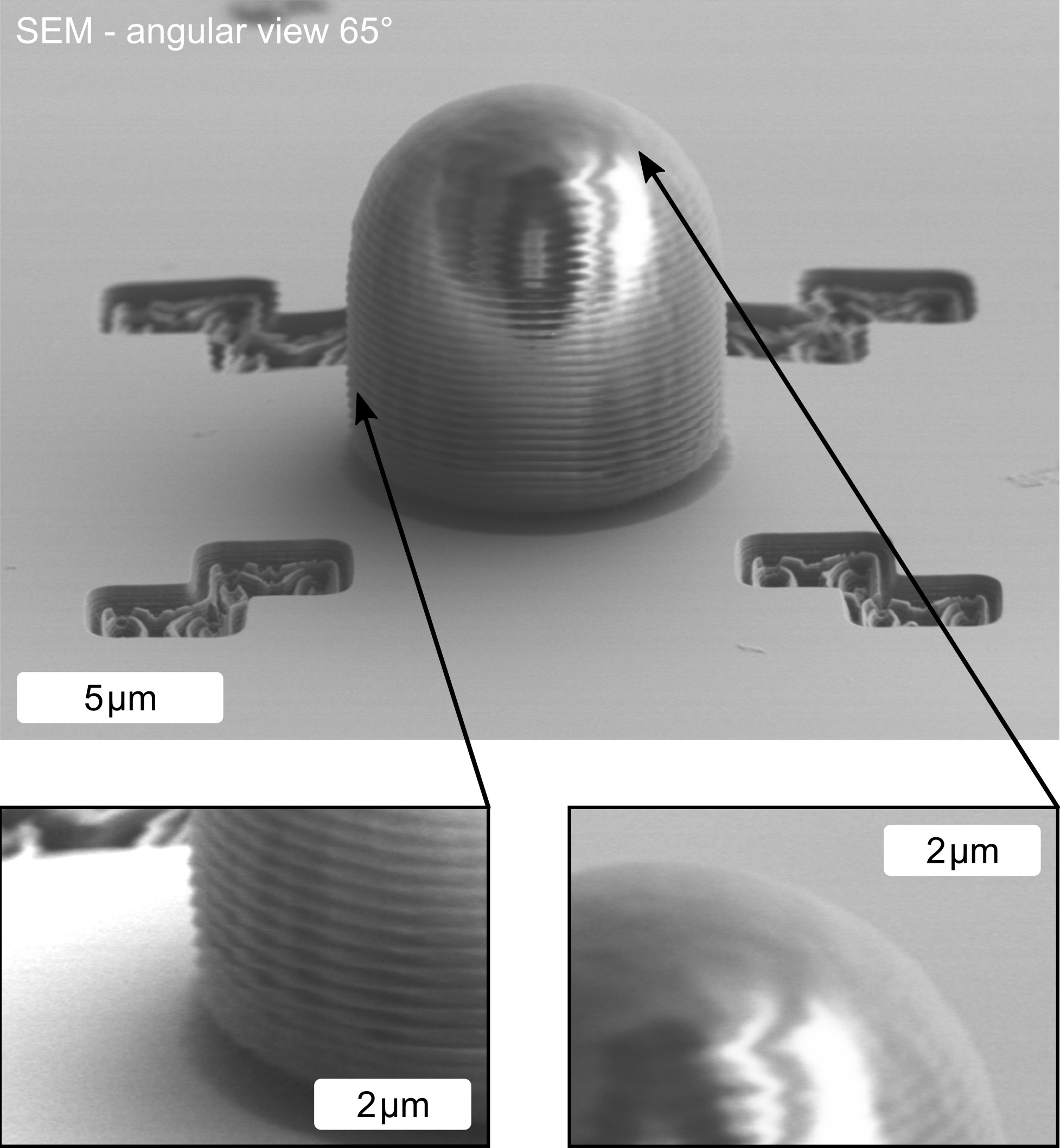}
	\caption{SEM angular view picture (\SI{65}{\degree} tilt) of the same W-SIL as depicted in the inset of \fig\ref{fig:weierstrasssilthesis}. Severe surface roughening is observable at steeper surfaces (lower left image) while curved surfaces show less roughness (lower right image). Additionally, the lower shape deviates from the nominal Weierstrass geometry.}
	\label{fig:weierstrasssilthesis65deg}
\end{figure}

According to the spectral comparison shown in \fig\ref{fig:weierstrasssilthesis}, a clear enhancement of all emission lines is observable, as well as the typical expected blue shift of around \SI{3}{\nm}.
Characterization before and after the SIL placement has been carried out in this case in a standard free-space \si{\micro}-PL setup with a detector collection NA of 0.45 under CW excitation with a He-Ne laser. 
Performing the data analysis on the X and CX emission lines as in the previous section \ref{section:h-SILs} resulted in a measured PL-intensity ratio of $\tilde{\eta}=3.85\pm0.45$. This value deviates quite much from the theoretically expected enhancement factor of around 6 for a detector collection NA of 0.45.
However, this behavior can be explained by looking at the investigated W-SIL shape and surface structure in \fig\ref{fig:weierstrasssilthesis65deg}. 
When investigating the lens shape via SEM in angular view (\SI{65}{\degree} tilted), it can be seen that the shape does not perfectly correspond to the Weierstrass geometry. Indeed, the lower part differs quite a lot from the ideal shape which causes losses due to refractions not folding the light into a small NAs. 
In addition, severe surface roughening is also visible in the lower left image in \fig\ref{fig:weierstrasssilthesis65deg} which is in the order of several tens of \si{\nm} up to \SI{100}{\nm}. This causes uncontrolled diffraction, which also limits the lens performance and explains the deviation from the theoretically expected value of around 6.
However, the apex of the lens seems to be quite smooth (lower right SEM image in \fig\ref{fig:weierstrasssilthesis65deg}).
Assuming the same placement accuracy as for the h-SILs this can also be one of the limiting factors since the Weierstrass geometry is supposed to be much more sensitive to the emitter position. 
Another W-SIL with the same shape as visible in \fig\ref{fig:weierstrasssilthesis65deg} was investigated and shows a PL-intensity enhancement of only $2.51\pm0.10$ which is assumed to occur due to an even larger lens displacement.

\subsection{Optical characterization of TIR-SILs}
\label{section:TIR_sils}
\begin{figure}[!h]
	\centering
	\includegraphics[,width=0.7\linewidth]{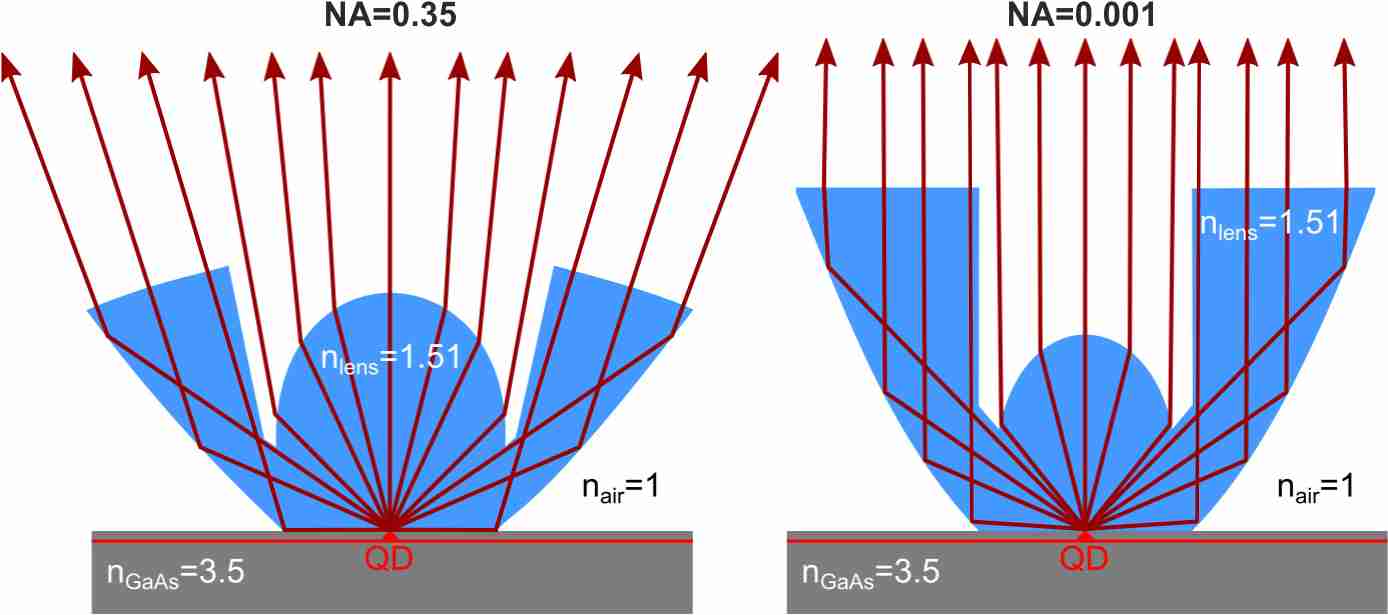}
	\caption{Cross sectional sketch of the working principle of a TIR-SIL which folds al the light leaving the semiconductor into a predefined NA. Reflections and refractions were calculated for this illustration with the commercially available ray tracing software OpticStudio. \textbf{(left)} TIR-SIL designed for $\mathrm{NA}=0.35$. \textbf{(right)} TIR-SIL designed for $\mathrm{NA}=0.001$. }
	\label{fig:tirsilsskizze}
\end{figure}
The W-SIL results showed that much higher values of collection efficiencies can be achieved with more complex lens geometries than an h-SIL and that care must be taken to avoid roughness or deviations from the optimal design.
For this reason, we developed an advanced SIL design referred to as TIR-SILs. This geometry makes use of the total internal reflection condition at the SIL-to-air interface. 
A cross sectional illustration which takes ray tracing calculations (with the commercially available software OpticStudio (\textit{ZEMAX})) into account, is shown in \fig\ref{fig:tirsilsskizze}. 

\begin{figure}[!h]
	\centering
	\includegraphics[]{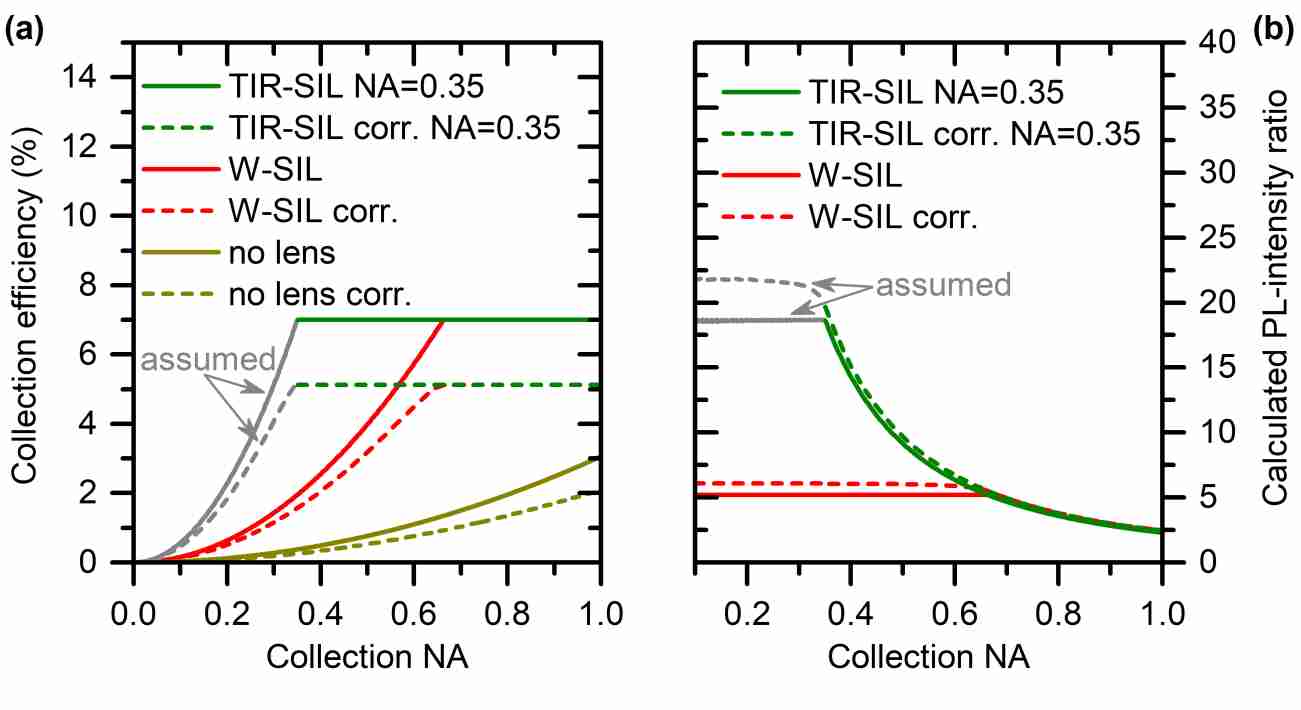}
	\caption{\textbf{(a)} Collection efficiencies for planar samples, W-SILs, and TIR-SILs (output NA of 0.35) versus the detector collection NA.
		\textbf{(b)} PL-intensity ratios for W-SILs and TIR-SILs (output NA of 0.35) versus the detector collection NA. The curves are obtained by division of the respective collection efficiency with the TIR-SIL by the planar one in (a).
		Solid curves represent ideal values, while dashed curves correct for one reflection at the present interfaces (labeled as ``corr.'').  
		The assumed continuation for lower NAs than 0.35 of the curves for the TIR-SIL geometry is based on the similarity of the inner Weierstrass-like lens shape and plotted in gray.}
	\label{fig:extraction-efficiencies-tir-sil}
\end{figure}
The TIR-SIL is composed of an inner aspheric lens and an outer parabolic reflector structure. The reflector makes use of the TIR condition at the lens-to-air interface as it can be seen in \fig\ref{fig:tirsilsskizze}.
By engineering both components, all the light can be folded into arbitrary output NAs.
Since total internal reflection is very sensitive to the angle of the incident light, a small lateral emitter misplacement is expected to affect the extraction in a more negative way compared to the h-SIL and W-SIL geometry. Thus, losses are induced in the form of light being refracted instead of fully reflected.
Choosing this TIR-SIL geometry can be interpreted in such a way that the kink in the efficiency curve of the W-SIL 
is shifted in the horizontal direction towards smaller NAs, depending on the lens design. 
This is illustrated (not calculated) in \fig\ref{fig:extraction-efficiencies-tir-sil} for a designed NA of 0.35.
\begin{figure}[!h]
	\centering
	\includegraphics[]{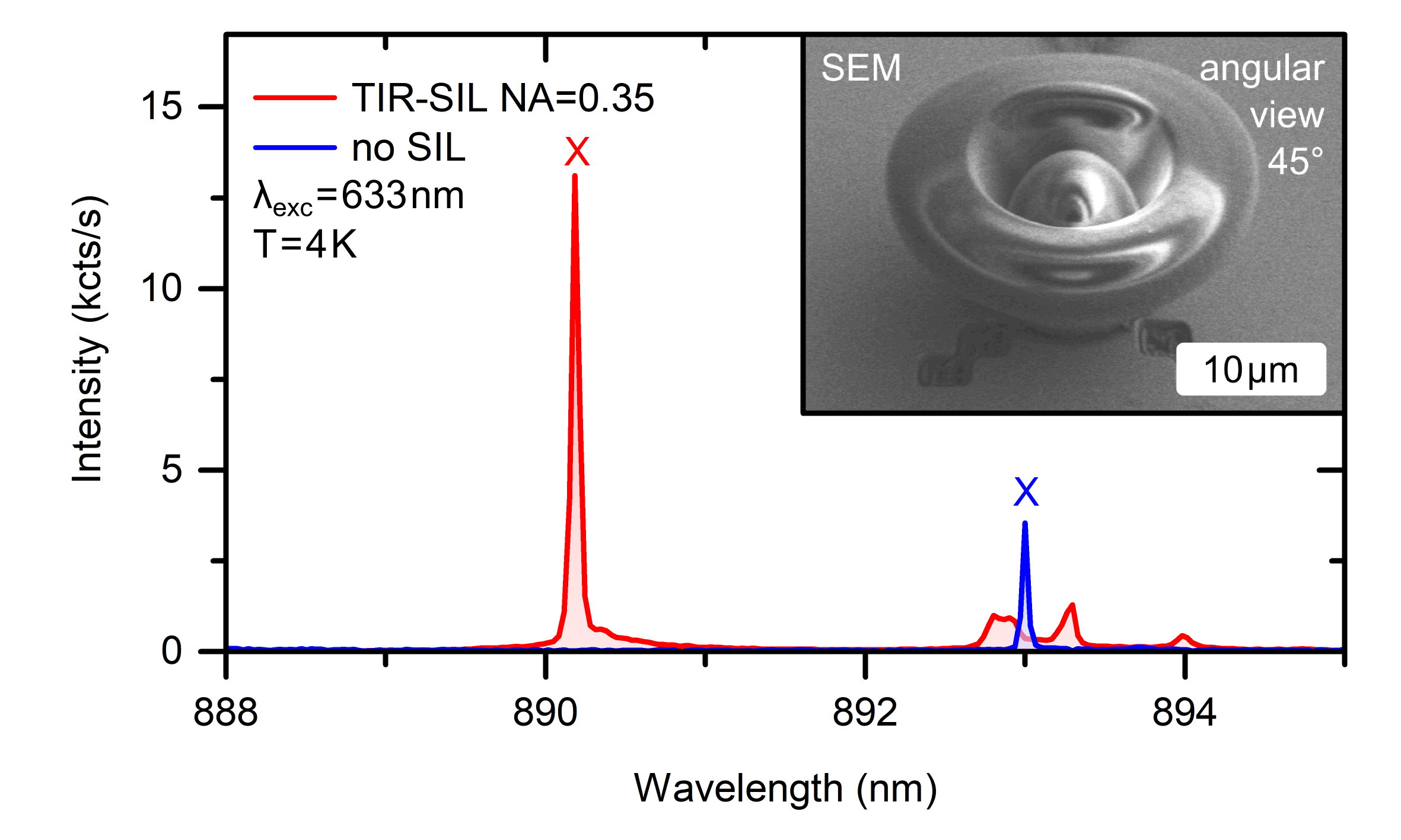}
	\caption{\si{\micro}-PL spectra of the same QD underneath a TIR-SIL designed for an NA of 0.35 with a bottom diameter of \SI{10}{\um} and without a lens. A blue shift of the main emission line of \SI{2.82}{\nm} is observable while the intensity is enhanced ba a factor of $\tilde{\eta}=5.61\pm0.14$.
		The inset shows a SEM angular view picture (\SI{45}{\degree} tilt) of the printed TIR-SIL designed for folding all the light leaving the \ce{GaAs} into an NA of 0.35.}
	\label{fig:tirsil035}
\end{figure}
The green sections of the curves can be calculated from the lens design. The gray sections can be justified based on the similarity of the inner Weierstrass-like lens shape.\\

In this work, two TIR-SIL geometries with different output NA designs were fabricated. Both are illustrated in \fig\ref{fig:tirsilsskizze} ((left) $\mathrm{NA}=0.35$ and (right) $\mathrm{NA}=0.001$).
Investigation of this TIR-SIL (see \fig\ref{fig:tirsil035}) with an NA of 0.35 results in a blue shift of \SI{2.82}{\nm} (\SI{4.4}{\milli\electronvolt}). This lies within the typical range for QDs underneath 3D printed objects with contact surfaces in this size. 
With the same data evaluation as before, a PL-intensity enhancement of $\tilde{\eta}=5.61\pm0.14$ (collection objective $\mathrm{NA}=0.45$) could be determined. 
Theoretical values of $\tilde{\eta}_\mathrm{ideal}=11.30$ and $\tilde{\eta}_\mathrm{corr}=11.96$ for a detector collection NA of 0.45 are extracted from the curves in \fig\ref{fig:extraction-efficiencies-tir-sil}.
\begin{figure}[!h]
	\centering
	\includegraphics[]{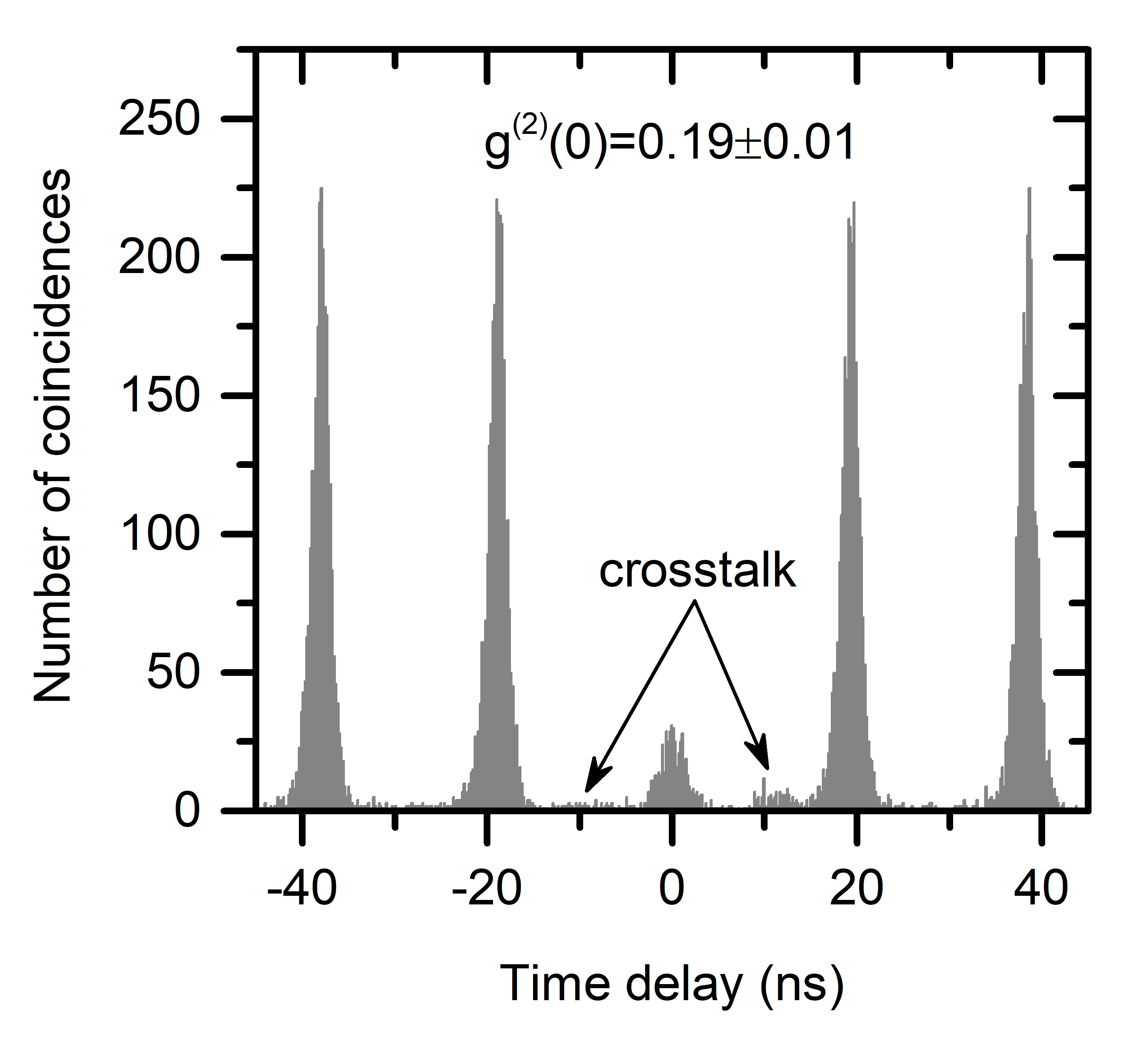}
	\caption{Autocorrelation measurement of the main emission line of \fig\ref{fig:tirsil035} under pulsed excitation at \SI{650}{\nm}. 
		For the $g^{(2)}(0)$ evaluation, a binning of \SI{7}{\ns} was chosen resulting in a raw $g^{(2)}(0)=0.19\pm0.01$. The recorded events between the central peak and the neighboring peaks occur due to cross-talk between both used APDs.}
	\label{fig:g2tir035saturation}
\end{figure}
The rather big deviation of the measured value from the theoretically expected one can be explained with the surface roughness of the printed material (see \fig\ref{fig:weierstrasssilthesis65deg}) and most probably due to a non-ideal placement of the lens.
In addition, the same QD below the lens has now been used to perform an autocorrelation measurement under pulsed excitation at \SI{650}{\nm} on the brightest line in saturation. This resulted in the histogram displayed in \fig\ref{fig:g2tir035saturation}. 
A binning of \SI{7}{\nano\second} gives a $g^{(2)}(0)$-value of $0.19\pm0.01$, which is a clear indicator of single-photon emission. 
Although the QD experiences massive strain during the cooling down after the 3D printing, the single-photon character of the emission is maintained.\\

\begin{figure}[!h]
	\centering
	\includegraphics[]{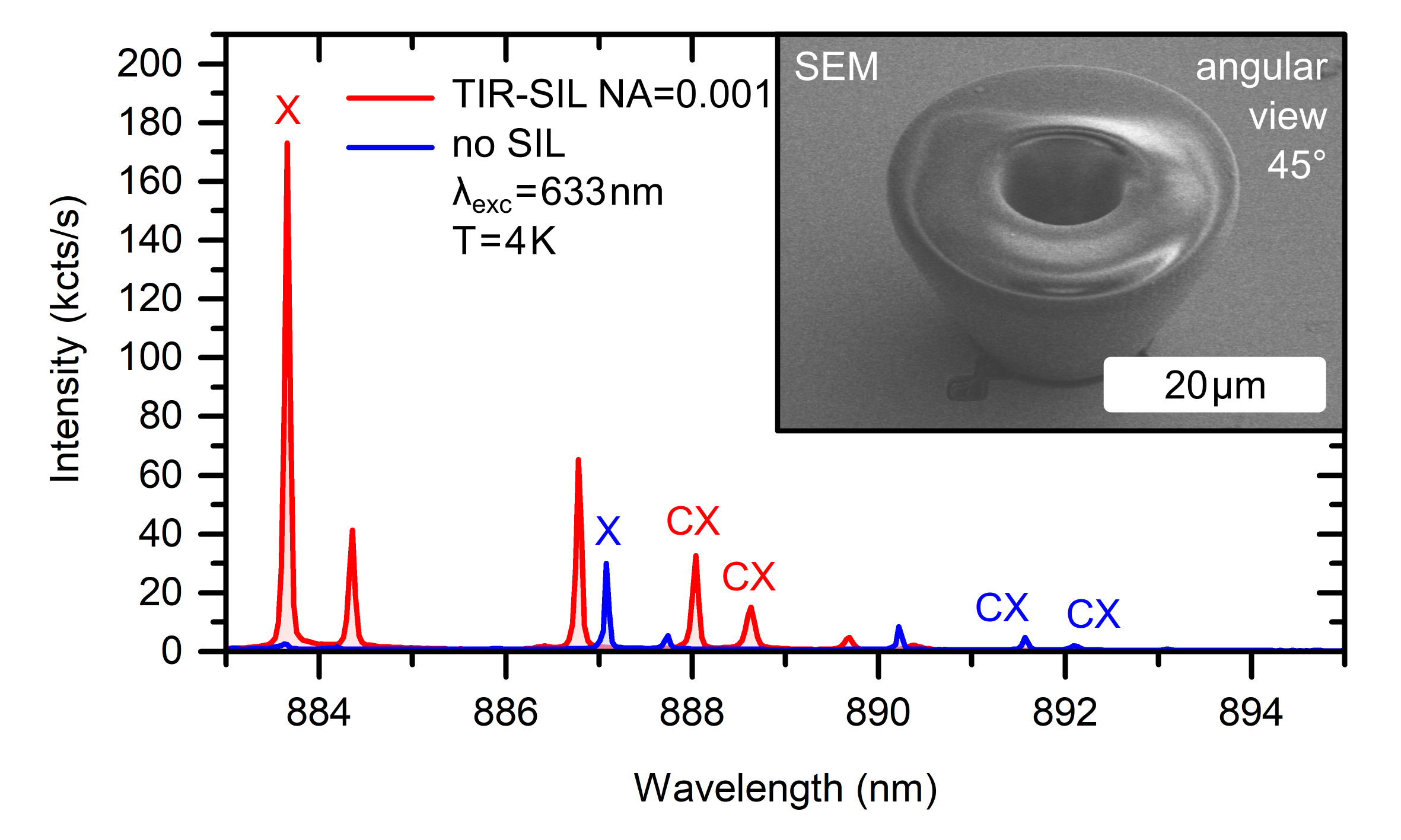}
	\caption{\si{\micro}-PL spectra of the same QD underneath a TIR-SIL designed for a NA of 0.001 with a bottom diameter of \SI{10}{\um} and without a SIL. Emission characteristics were identified prior to the intensity enhancement evaluation via power-dependent measurements. The inset shows an SEM angular view picture (\SI{45}{\degree} tilt) of the printed TIR-SIL designed for folding all the light leaving the \ce{GaAs} into an NA of 0.001.}
	\label{fig:tirsilsthesis}
\end{figure}
In order to gather further statistics for the TIR-SILs, several SILs folding the light into a NA of 0.001 were 3D printed while being aligned on the pre-selected QD positions. 
\fig\ref{fig:tirsilsthesis} displays two spectra of the same QD before and after the 3D printing. 
Again, a blue shift of \SI{3.43}{\nm} (\SI{5.42}{\milli\electronvolt}), which is comparable with previously achieved results, can be observed. 
A PL-intensity ratio of $8.33\pm0.39$ can be deduced for this lens. 
The inset shows an angular SEM picture (\SI{45}{\degree} tilt) of the lens design. It is aligned here exemplary on etched markers, even though the spectrum belongs to a QD marked with metal markers.
By the deposition of metal markers, distortion of the alignment laser beam at the marker edges can be avoided. Therefore, an improvement in the alignment accuracy of the 3D direct laser writing machine is expected. 
5 TIR-SILs were fabricated in total, only one of them with etched markers. 
\begin{figure}[!h]
	\centering
	\includegraphics[]{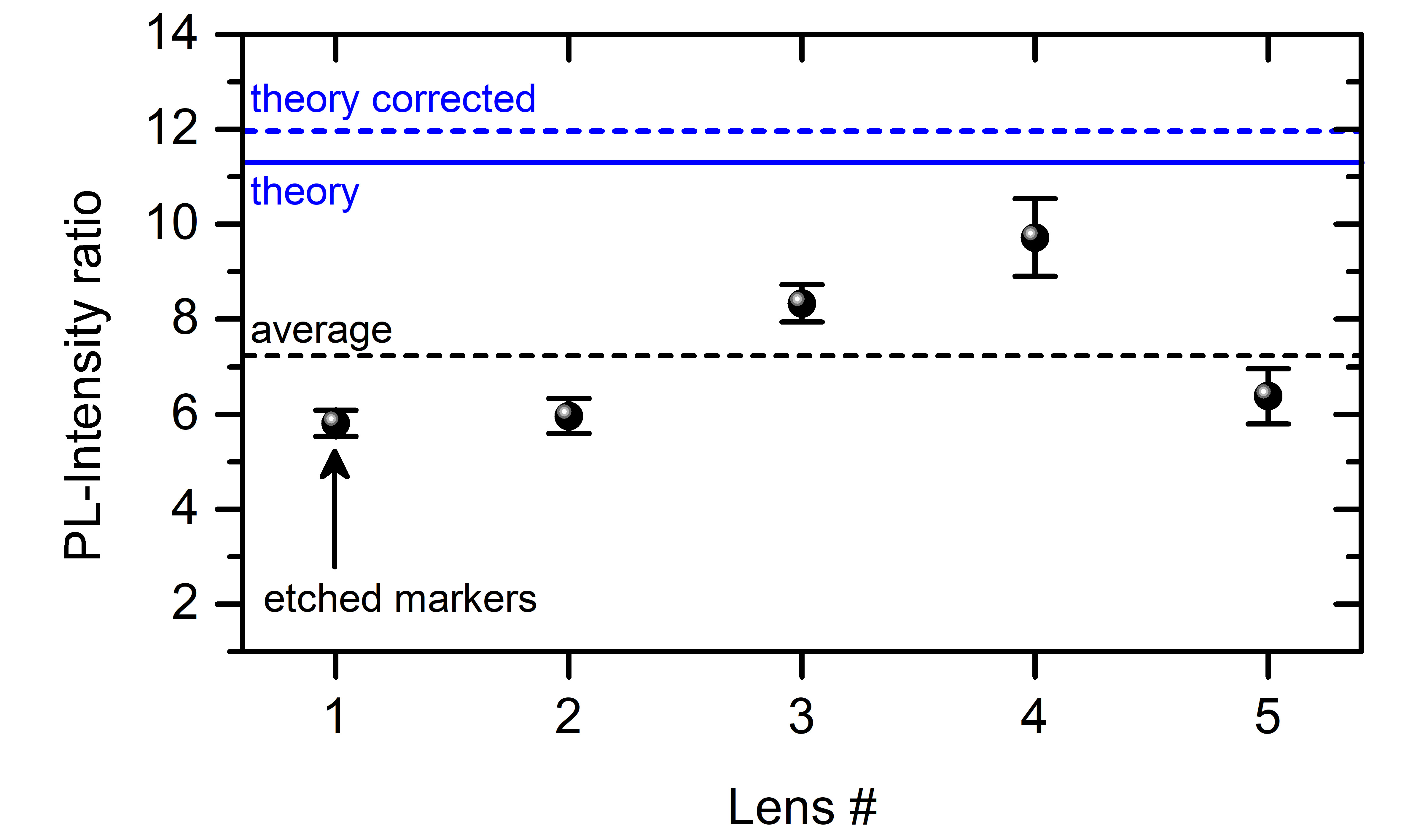}
	\caption{PL-intensity ratios for a total of 5 printed TIR-SILs, which fold all light into a NA of 0.001.
		Lenses yield an enhancement between about 6 and 10.
		The lateral position of the first lens was determined using etched markers, while all other lenses were positioned using metallic markers.}
	\label{fig:statisticstir0k001}
\end{figure}
Analysis of this lens also resulted in the lowest enhancement factor (see \fig\ref{fig:statisticstir0k001}) since it is more difficult to align the laser of the 3D printing machine on these markers.
For the other four lenses, markers in \ce{Cr}/\ce{Au} were deposited. 
A maximum PL-intensity ratio of $9.72\pm0.82$ was observed here, underlining the use of metal markers and the sensitivity of the enhancement factor on the lateral lens placement accuracy.
On average, a PL-intensity enhancement ratio for this lens geometry of $\tilde{\eta}= 7.24\pm0.49$ could be determined which is marked as the black dashed line in \fig\ref{fig:statisticstir0k001}. Respective calculated enhancement factors are illustrated in blue.\\

Tab.\,\ref{tab:overviewenhancement} gives a final overview on the theoretically expected and averaged measured PL enhancement factors for each fabricated lens geometry. For future applications, the TIR-SIL designed for a output NA of 0.001 offers the most promising results, especially when the 3D printing machine is aligned on metal markers.
\begin{table}[h]
	\centering
	\begin{tabular}{|c|c|c|}
		\hline
		\textbf{Lens type} & \textbf{\begin{tabular}[c]{@{}c@{}}Theoretical PL\\ enhancement\end{tabular}} & \textbf{\begin{tabular}[c]{@{}c@{}}Measured PL\\ enhancement\end{tabular}} \\ \hline
		h-SIL              &                     2.65                                                          &                                     $2.09\pm0.23$                                       \\ \hline
		W-SIL              &               6.02                                                                &                                         $3.18\pm0.28$                                   \\ \hline
		TIR 0.35 NA        &               11.96                                                                &                           $5.61\pm0.14$                                                 \\ \hline
		TIR 0.001 NA       &                   11.96                                                            &                 $7.24\pm0.49$                                                           \\ \hline
	\end{tabular}
\caption{Overview on theoretically expected and average measured PL enhancement factors for each fabricated lens type.}
\label{tab:overviewenhancement}
\end{table}

\subsection{Deterministic fiber incoupling}
\begin{figure}[!h]
	\centering
	\includegraphics[]{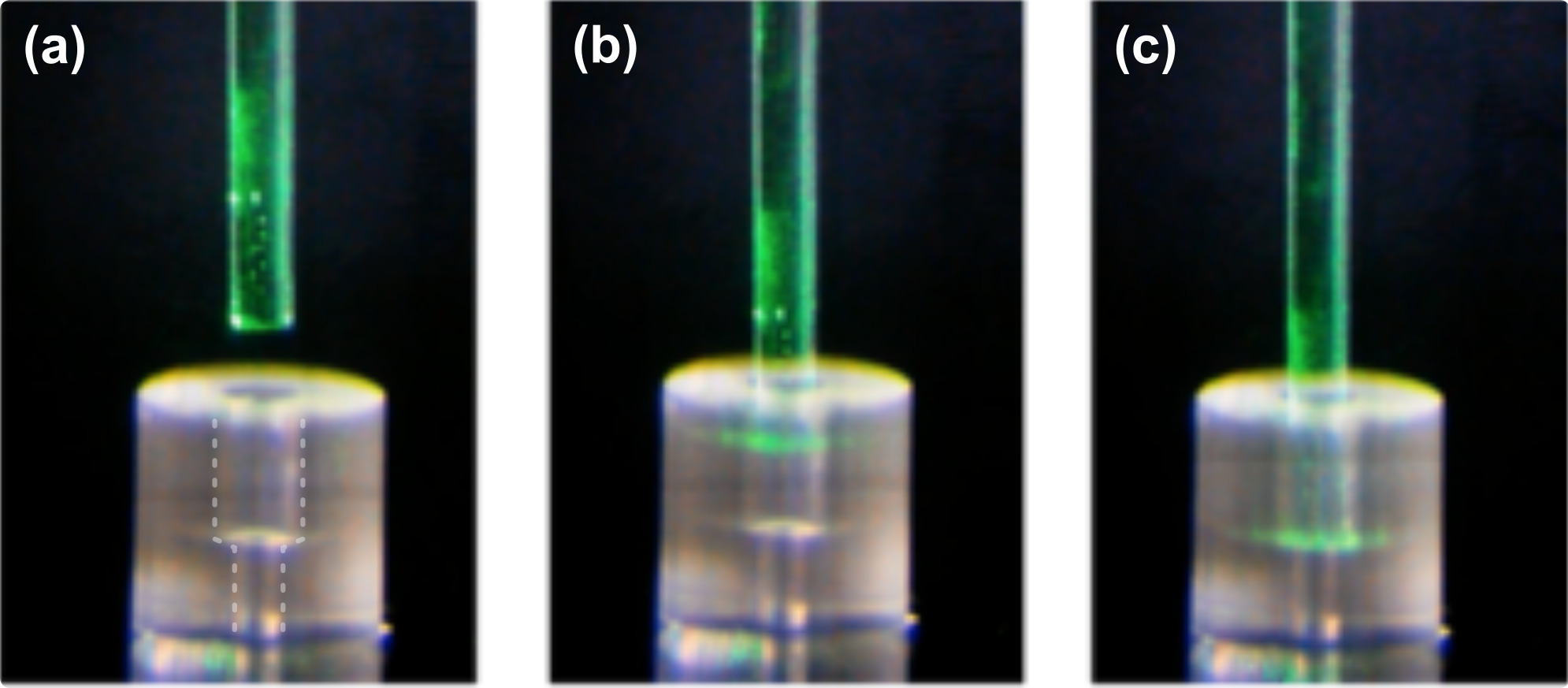}
	\caption{Different stages of the fiber mating process recorded as a sequence of microscope pictures. \textbf{(a)} The fiber is aligned on the opening hole of the chuck. \textbf{(b)} Intermediate snapshot of the mating procedure while the fiber is moved towards the stopping level.
		\textbf{(c)} Completed fiber mating. The fiber is stopped via the step indicated in (a) by the dashed white lines and is ready for being fixed with epoxy glue.}
	\label{fig:fibermating}
\end{figure}
\begin{figure}[!h]
	\centering
	\includegraphics[draft=false]{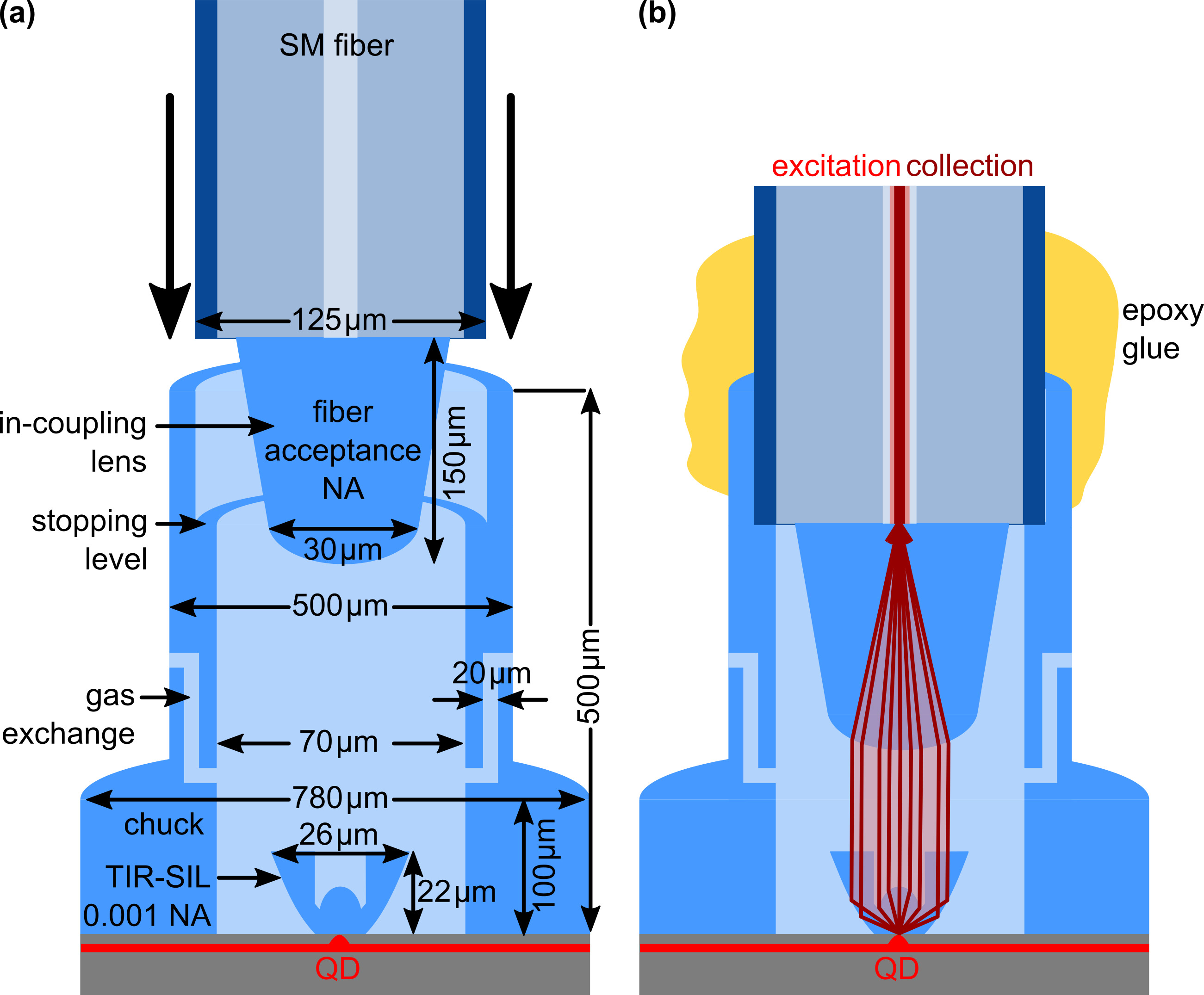}
	\caption{Schematic of the fiber chuck design. \textbf{(a)} A TIR-SIL with an NA of 0.001 is printed, determinstically aligned on the QD position. After the characterization of the printed lens, the big tube-like chuck is fabricated, being aligned on this lens. On the fiber tip, another lens is printed for coupling the modified emission into the fiber core. The modified fiber is then inserted into the chuck. Typical dimensions of the 3D printed parts are labeled respectively. \textbf{(b)} Epoxy is used for fixing the fiber position. Excitation and collection of the QD are carried out via the same fiber, here illustrated as light red (excitation) and dark red (collection).}
	\label{fig:fiberchuck}
\end{figure}
\begin{figure}[!h]
	\centering
	\includegraphics[]{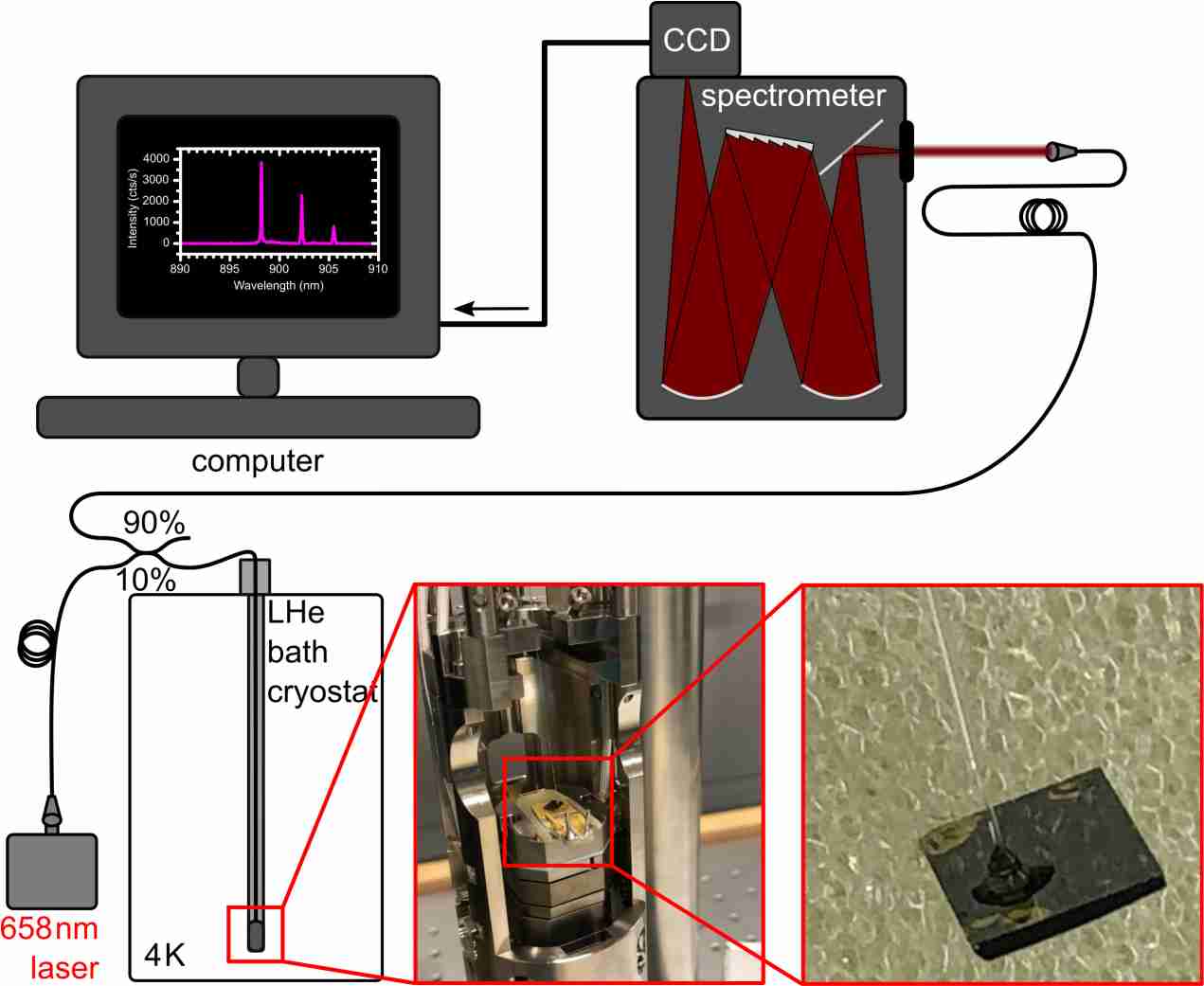}
	\caption{Setup configuration for the characterization of the standalone fiber-coupled device. The optics head is not needed in this case. Excitation takes place via a fiber coupled laser diode at \SI{658}{\nm}, sent through a 90:10 beam splitter and a fiber troughport connected to the sample. Light, emitted by the QD is guided via the \SI{90}{\percent} channel to the spectrometer and recorded via a CCD on the computer.}
	\label{fig:fiberchucksetup}
\end{figure}
\begin{figure}[!h]
	\centering
	\includegraphics[]{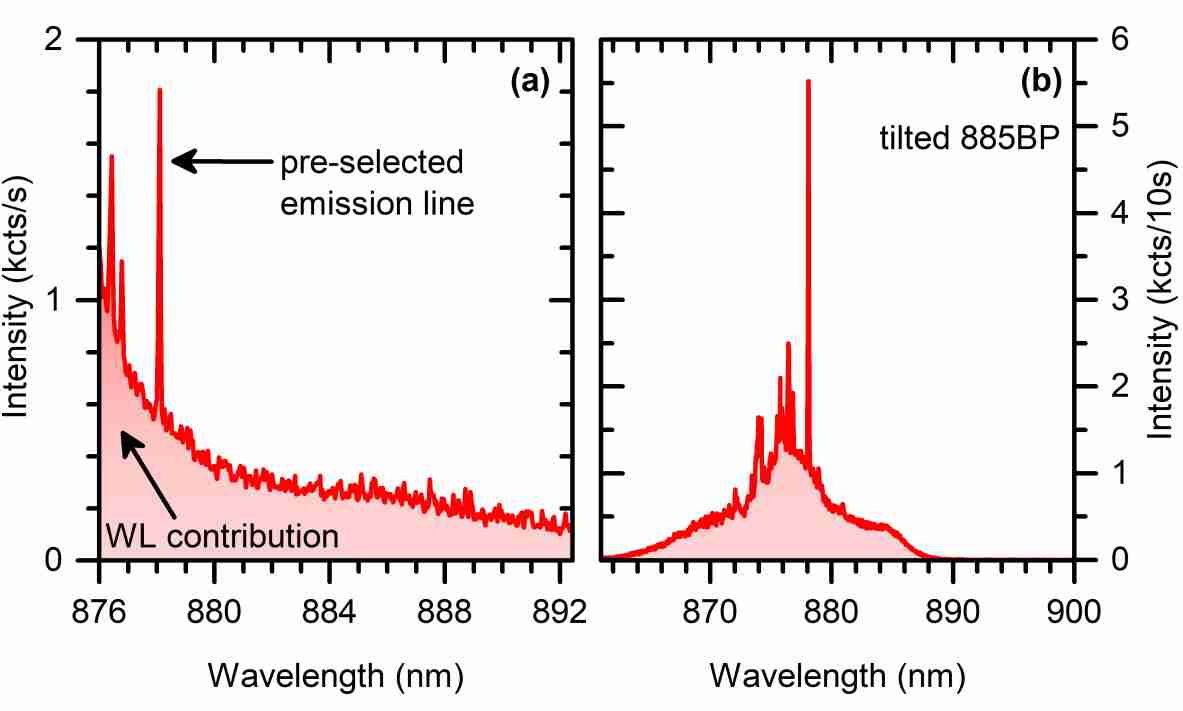}
	\caption{\textbf{(a)} Unfiltered PL signal of the standalone QD device. \textbf{(b)} Spectrum filtered with a band-pass filter which is designed for $\SI{885}{\nm}\pm\SI{12.5}{\nm}$. Tilting the filter shifts the wavelength window down to lower wavelengths. The big printed chuck induces additional strain on the QD causing a further blue shift of the emission to the edge of the wetting layer. Both spectra were taken in saturation. The QD with the spectrum in \fig\ref{fig:tirsilsthesis} was used for the fiber mating.}
	\label{fig:fibermatingspectrum}
\end{figure}

After the evaluation of the different micro-lens designs, we now focus on the fabrication of a single mode fiber-coupled standalone single-photon device based on 3D direct laser writing. For this purpose we develop a design for a 3D printed fiber chuck.
The working principle is depicted in \fig\ref{fig:fibermating}. \fig\ref{fig:fibermating}(a) illustrates the chuck and the fiber before the mating takes place. The mating is performed with a manual XYZ-flexure stage under a microscope. 
The guideway is highlighted by the dashed lines in \fig\ref{fig:fibermating}(a). 
The narrower diameter at the lower level ensures a fixed vertical alignment, while the higher level tube serves as the restriction in the lateral degree of freedom.
After the lateral alignment, the fiber is inserted into the chuck, which can be seen in the intermediate microscope picture \fig\ref{fig:fibermating}(b). Once the fiber cannot be moved any further (\fig \ref{fig:fibermating}c), the mating is complete and epoxy glue is applied to fix the fiber to the 3D printed structure.
\begin{figure}[!h]
	\centering
	\includegraphics[]{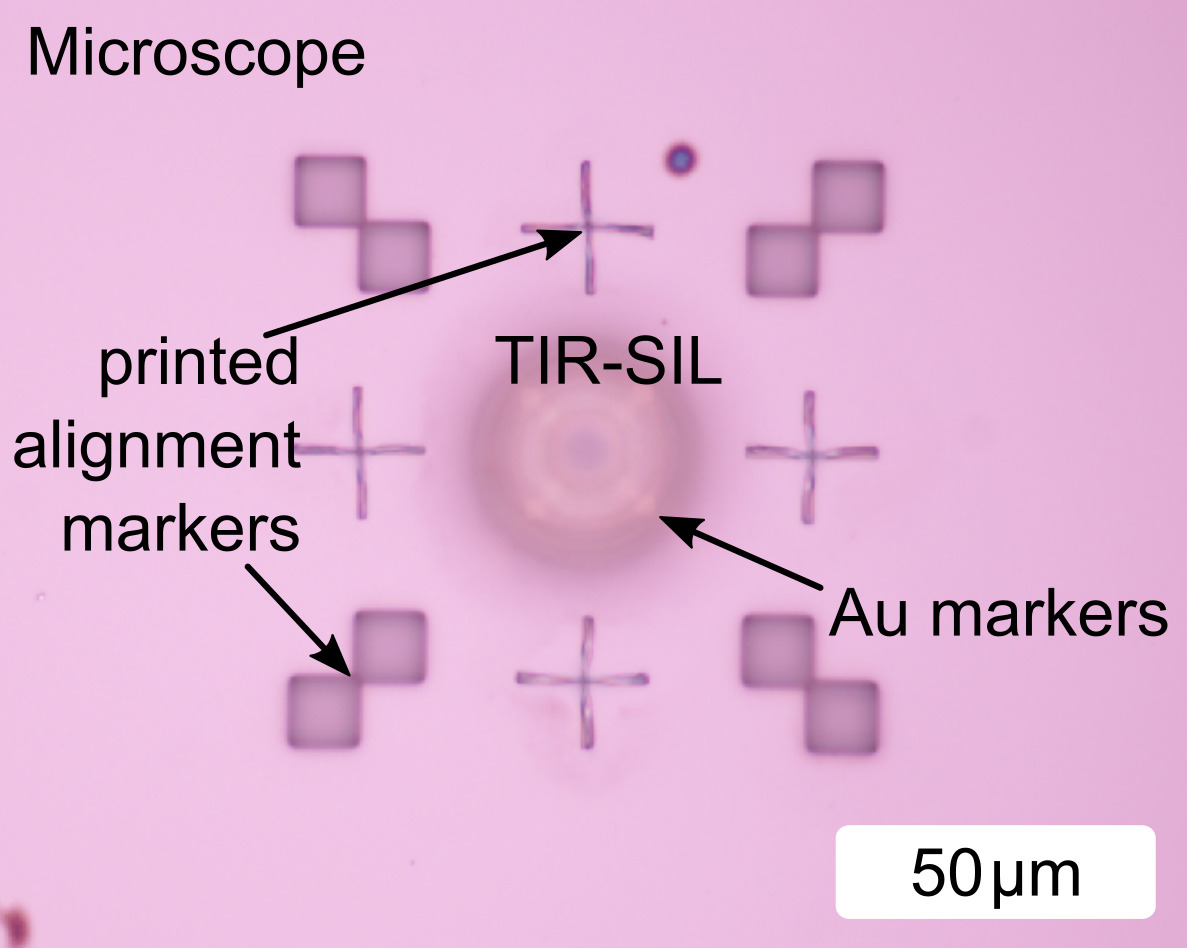}
	\caption{Microscope image of a TIR-SIL designed for an NA of 0.001. Alignment markers were printed in the same process step with the scope of later alignment for the fiber chuck printing.
		Deposited Au markers are still partly visible underneath the TIR-SIL.}
	\label{fig:fiberchuckbefore}
\end{figure}
The entire structure is built on the precise knowledge on the location of a QD, provided by the deposition of metal markers, and the 3D direct laser writing of a TIR-SIL. 
In order to properly align the large chuck to the lens position, the step in which the TIR-SIL is printed has to be slightly adapted.
The lateral lens dimensions are so large that the metal markers are no longer visible after 3D printing. Therefore, the fabrication of the lens is supplemented by simultaneous 3D printing of another set of large standard markers at a distance of a few \si{\um}. An exemplary microscope image of a TIR-SIL with the 3D printed markers can be seen in \fig\ref{fig:fiberchuckbefore}.
The chuck printing step is then aligned to the 3D printed markers and
a conical incoupling lens is additionally printed on the fiber tip which is sketched in \fig\ref{fig:fiberchuck}(a)).
It is designed for refracting the incident QD emission into the acceptance NA of the single-mode fiber (SM-780HP, NA = 0.14). 
When the alignment is finished, the fiber and the chuck are fixed by epoxy glue as sketched in \fig\ref{fig:fiberchuck}(b). 
In this experiment, the epoxy glue unfortunately also covered the gas exchange channels (sketched in \fig\ref{fig:fiberchuck}). 
They were implemented with the idea of avoiding humidity and air freezing inside the device during the cooling cycle.
Air inside the chuck is supposed to be evacuated when the sample is placed into a cryostat.
Since this was not possible in this experiment round, it limited the device performance.
For the sample characterization, the fiber-coupled QD SPS is mounted in the deterministic lithography setup as can be seen in \fig\ref{fig:fiberchucksetup}. 
The red box indicated the mounted sample in more detail.
On the bottom right, the standalone device is displayed completely covered in epoxy which provides robustness in terms of mechanical stability when being mounted.
It is then glued to the sample holder and placed inside the setup as in the mid bottom inset in \fig\ref{fig:fiberchucksetup}.
The installed fiber throughport at the top of the setup ensures the fiber connection from the red excitation laser to the sample and back to the spectrometer via a 90:10 beam splitter. 
All fibers are connected in a way that only \SI{10}{\percent} of the QD emission is lost at the beam splitter.
After cooling down to \SI{4}{\kelvin}, the QD is excited through the fibers with a laser module emitting at \SI{658}{\nm} and the emission is guided back to the spectrometer.
Due to its high brightness and clean spectrum, the QD investigated in \fig\ref{fig:tirsilsthesis} was chosen to be the ideal candidate for the fiber mating experiment.
The investigation results in the saturated spectra shown in \fig\ref{fig:fibermatingspectrum}(a) and (b).
It is directly visible from the left spectrum that due to the big 3D printed chuck even more strain is induced as a result of the lens material contraction at low-temperatures.

\section{Discussion}

The brightest emission line is located in \fig\ref{fig:fibermatingspectrum}(a) at \SI{878.12}{\nm}.
Unfortunately, the spectrum could not be directly reproduced because of the strain being so large that the QD emission shifted to the tails of the wetting layer. The further blue shift could only be roughly estimated (\SI{5.55}{\nm} which equals \SI{8.9}{\milli\electronvolt}) by investigation of the brightest observable emission line. 
It can be concluded that the lens and the chuck induce a blue shift on pre-selected QDs of around \SI{9}{\nm} in this emission wavelength range (in energy approximately \SI{13}{\milli\electronvolt}).
In analogy to previous estimations of the present stress on the QD, a strain of around \SI{468}{\mega\pascal} can be derived \cite{ding2010,trotta2015}.
For further experiments and applications, this shift has to be considered in the QD pre-selection step by selecting QDs with emission wavelengths far away from the wetting layer.
In order to partially suppress the wetting layer contribution when acquiring a spectrum where the emission line is spectrally in the center of investigation, a band-pass filter (885BP) was placed inside the detection path in front of the spectrometer entrance. 
By tilting this filter, the window of transmission could be shifted towards shorter wavelengths. This suppresses the large contribution on the lower wavelength side from the wetting layer and results in the spectrum visible in \fig\ref{fig:fibermatingspectrum}(b) (integration time \SI{10}{\second}).

\begin{figure}[!h]
	\centering
	\includegraphics[width=0.8\linewidth]{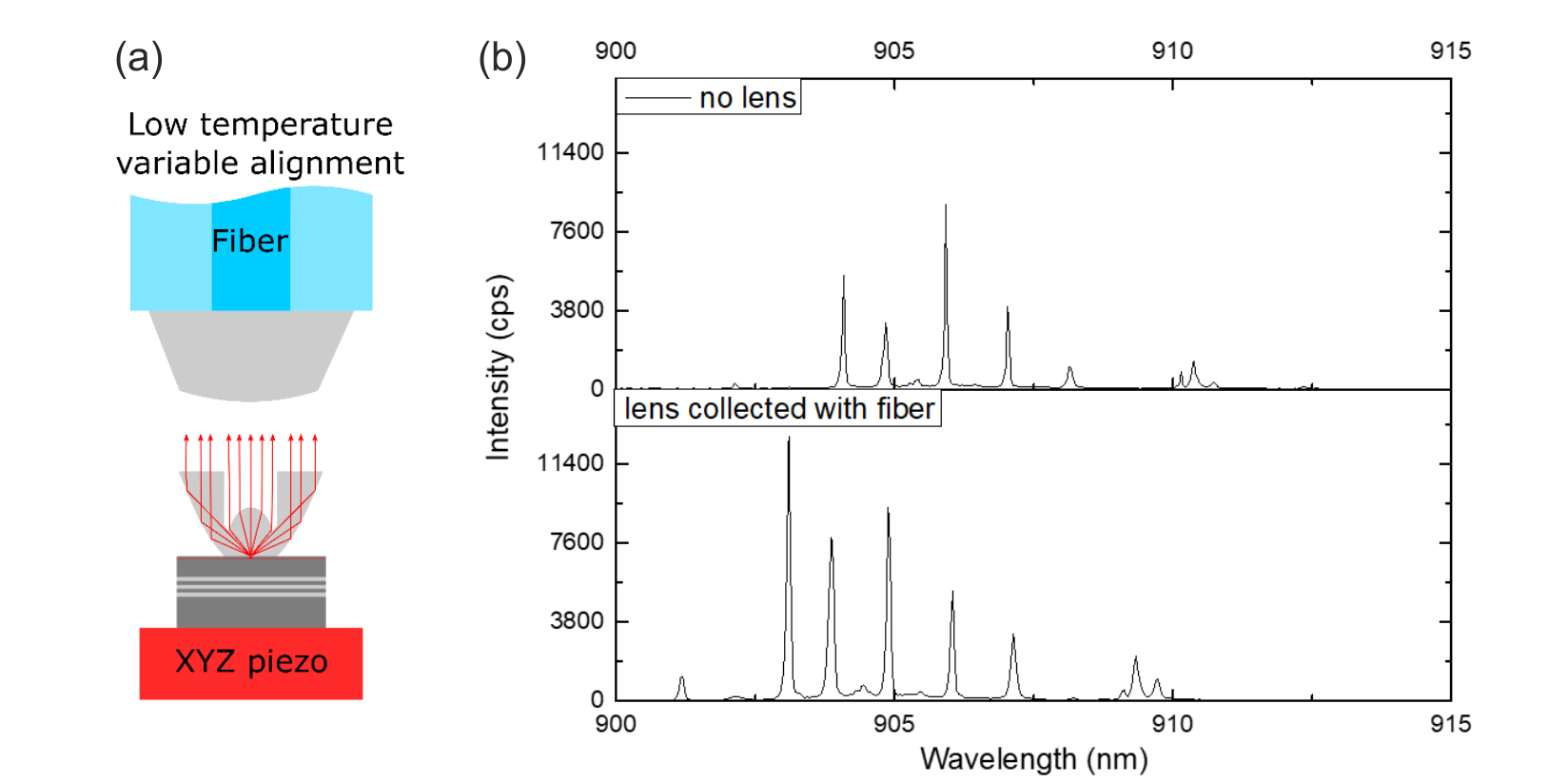}
	\caption{\textbf{(a)} Schematic illustration of the measurement principle used to validate the incoupling efficiency. The sample is mounted onto a high precision xyz piezo stage and can thus be moved in all three dimensions in relation to the fiber.  \textbf{(b)} Spectrum of utilized QD before the lens printing (top) and maximum fiber coupled signal (bottom). Spectra are saturated upon above band gap pumping ($\approx$ 660 nm). Figure reproduced from the supplement of \cite{bremer2020}.}
	\label{fig:no_chuck}
\end{figure}

Comparing the integrated intensity of the coupled QD light from \fig\ref{fig:fibermatingspectrum}(a)  with the integrated intensity before the chuck fabrication (\fig\ref{fig:tirsilsthesis}) resulted in a total coupling efficiency of around $1.05\pm\SI{0.18}{\percent}$. One factor that might inhibit the incoupling efficiency is a non-perfect alignment between fiber lens and TIR-SIL, possibly induced by the cooling process. Due to the circular symmetric chuck design, the effect of material contraction when cooling down is not expected to induce displacements of the components in the lateral direction. Nevertheless, a change in the vertical spacing seems likely. Additionally, the material contraction of the optical elements, as well as the thermo-optical effect have to be taken into account, as they will influence the focal length of the fiber lens. Aside from these issues, the surface roughness of the lenses or  a displacement of the incoupling lens with respect to the single mode fiber core might contribute to the losses. 
In order to gain a better understanding of the limiting factors, we validate the incoupling efficiency achievable with the current lens design, by use of another QD sample without a chuck and perform an in-situ alignment of the fiber and TIR-SIL. As illustrated in \fig\ref{fig:no_chuck}(a), the QD sample can be moved in relation to the fiber in a high stability cryostat bath, while monitoring the incoupling signal. This way, we eliminate the displacement between fiber and TIR-SIL by aligning to the maximum signal. \fig\ref{fig:no_chuck}(b) compares the QD spectra without any lens to the fiber coupled signal obtained by this method. Performing the same analysis that was used previously for the evaluation of the h-SILS, we find an incoupling efficiency of $26(\pm\SI{2}{\percent})$, corresponding to a increase by more than a factor of 20 compared to our first result. From this we conclude that a precise positioning of the fiber via the 3D printed chuck is crucial for the performance of our system. A fully integrated device using a glued fiber chuck yielding the same efficiency (corresponding to maximum single-photon rate of \SI{1.5}{\mega counts / s}) was realized recently in \cite{bremer2020}. In the same work, successful Hanbury-Brown and Twiss measurements were performed, that resulted in a $g^{(2)}(0)$-value of 0.13$\pm$0.05 \cite{bremer2020}, confirming the single-photon nature of the fiber coupled emission. In the future, incoupling efficiencies might be improved further by addressing the issues mentioned above, like the influence of the thermo-optical effect, or the positioning of the lens on the fiber.

\section*{Conclusion}

In conclusion, we successfully developed a reproducible method to enhance the collection efficiency in single QD spectroscopy by the combination of deterministic, optical low-temperature lithography with femtosecond 3D direct laser writing. 
Here, the 3D printing machine was either aligned on etched markers or deposited metal markers which proved themselves as the better choice in terms of alignment procedure.
Different lens shapes were investigated experimentally and compared to calculations.
For all lens geometries, an increase in collection efficiency was confirmed.
The simplest geometry, namely the hemispheric SIL, resulted in an intensity enhancement of around 2.1, which is in good agreement with performed calculations.
Furthermore, the h-SILs seem to be relatively tolerant to displacement errors with respect to the QD position. 
Despite that, a spectral blue shift due to the contraction of the lens material and thus induced strain on the QD when cooled down to \SI{4}{\kelvin} has to be considered if a specific target emission wavelength is desired. 
Moreover, the lenses result in a better focusing of the pump laser, which affects the pumping condition, eventually modifying the ratio between different emission lines of the same QD.
In addition to that, an increased signal-to-noise ratio was observed pushing the localization accuracy of a QD from \SI{2}{\nm} down to below \SI{1}{\nm}.
A further increase in collection efficiency is promised by the Weierstrass geometry which posesses a hyperhemispherical shape.
It offers a more significant modification of the emission,
redirecting all the light leaving the semiconductor into a detector collection NA of around 0.66.
Enhancement factors of up to approximately 3.9 were observable
in combination with a typical spectral blue shift. The deviation from the predicted enhancement value can be attributed to a non-ideal alignment of the 3D printer on the etched markers and to a non-perfect lens shape and lens surface roughness which cause uncontrolled diffraction at the lens-to-air interface.
Further improvement has been achieved by the development of the TIR-SIL geometry which consists of a centered aspheric lens and a radial symmetric reflector structure, making use of the total internal reflection condition.
This approach has proven to fold the light in even smaller and customizable NAs than achievable for the Weierstrass geometry.
Theoretical estimates of the enhancement factor are also above the achieved experimental value. 
Since this lens type is relying on the total internal reflection condition, its placement accuracy becomes even more critical than before. Due to that, only one measured value resulted to be close to the expected theory.
In general, the TIR-SIL reliably provides a PL-intensity ratio between 6 and 10, also in combination with the overall spectral blue shift of around \SI{5}{\milli\electronvolt}.
To summarize, a reliable process was developed to efficiently enhance the collected QD light by a factor of up to 10. 
Using the TIR-SIL folding all the light into an NA of 0.001, the idea of a fiber coupled standalone quantum dot device was partially realized. 
Due to the large contact area of the fiber chuck and the sample, additional strain was induced leading to an additional blue shift of the overall emission. As a consequence, the spectrum of the one and only device overlapped with the wetting layer. 
By selection of QDs with a far higher wavelength, this issue can be avoided. The validation of the approach for fiber in-coupling, i.e. a QD provided with a TIR-SIL and a fiber with additional focusing lens, was performed employing a setup capable of precisely aligning the fiber with respect to the emitter. A value up to \SI{26}{\percent} was shown, opening the route to a stable stand-alone, fiber-coupled device \cite{bremer2020}.\\

In the future, we are going to optimize coupling and add anti-reflection coatings, trying to push the coupling efficiency to the theoretical limit. Furthermore, our technology can be combined with QD SPS based on circular Bragg gratings \cite{Ricket2019, Kolatschek2019}, NV-centers, defects, and other quantum emitters. Also, highly efficient combination with single quantum detectors should be feasible.

\section{Material and methods}

\subsection{Deterministic quantum dot fabrication}
\label{section:3Dprintingstructure}
\label{section:3DprintingFabrication}

\begin{figure}[!h]
	\centering
	\includegraphics[]{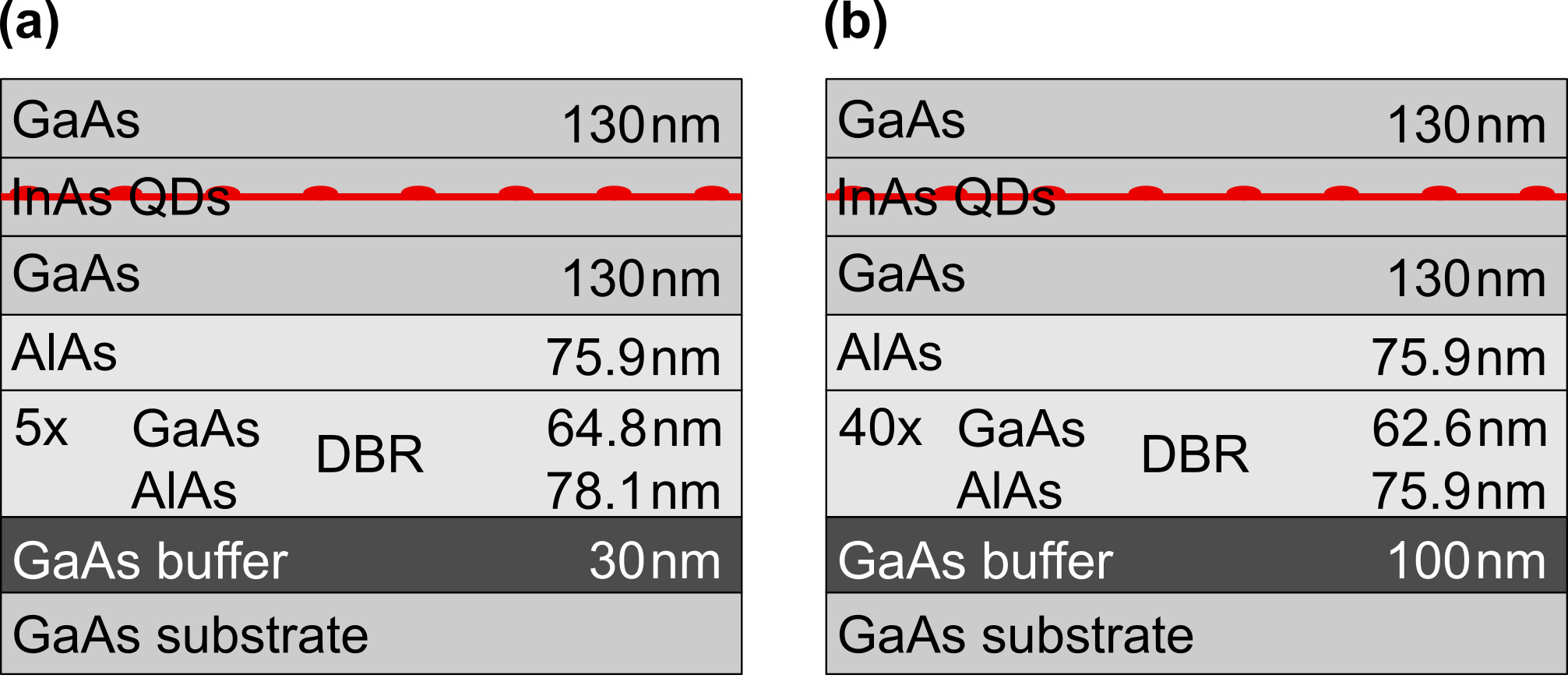}
	\caption{Schematic of the used sample structures. For both samples, a bottom DBR was grown to enhance the intrinsic QD brightness. \textbf{(a)} Used sample structure for the measurements in section \ref{section:h-SILs} and \ref{section:WSILs}. \textbf{(b)} Used sample structure for measurements in section \ref{section:TIR_sils}.} 	   \label{fig:samplestructure}
\end{figure}

\begin{figure}[!h]
	\centering
	\includegraphics[]{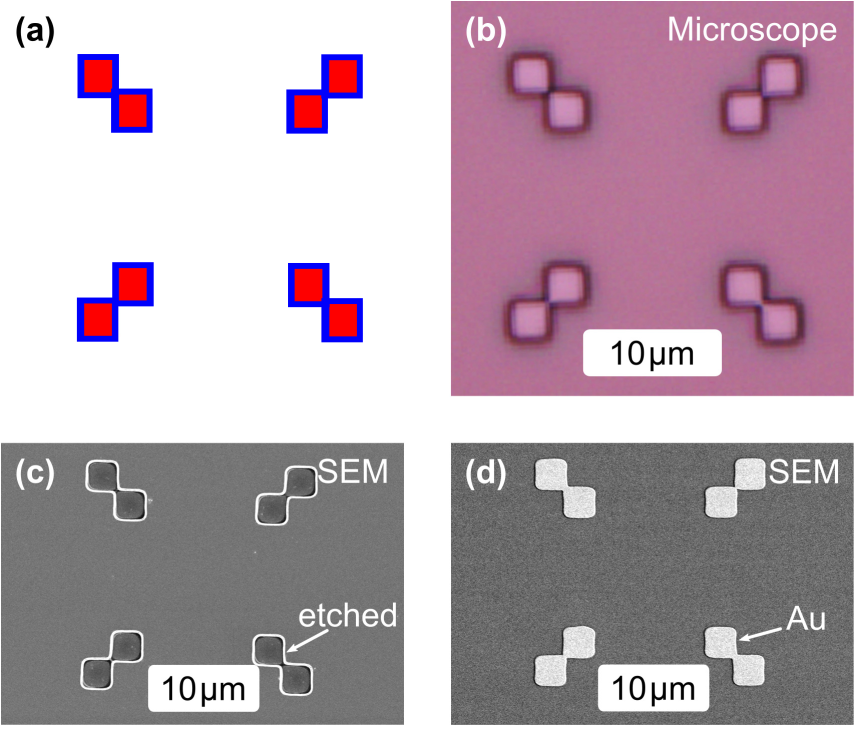}
	\caption{\textbf{(a)} Illustration of the marker geometry.  \textbf{(b)} Microscope picture of developed markers after positive lithography process. \textbf{(c)} SEM picture of markers transferred into the substrate with ICP-RIE etching. \textbf{(d)} SEM picture of \SI{50}{\nm} thick \ce{Au} markers after lift-off.}
	\label{fig:markerspng}
\end{figure}

Both  samples used in this work are schematically illustrated in \fig\ref{fig:samplestructure} and were grown via metalorganic vapour-phase epitaxy (MOVPE).
The overall structures are similar in the sense that both are based on the growth of a bottom AlAs/GaAs distributed Bragg reflector (DBR) structure below the InAs QD layer to enhance the emission brightness before the 3D printing of SILs. For details on the sample growth we refer to Refs. \cite{sartison2017, Herzog2018}.
Since we concentrate on PL-intensity enhancement and not on measuring directly the collection efficiency, the number of DBR-pairs is actually not decisive. 
At first, the sample is spin-coated with photoresists. It is then mounted with Fixogum, a silicon based glue, onto a sample holder  and  placed onto a piezo stack underneath a specially designed
low-temperature microscope objective with a numerical aperture of
NA = 0.8. The sample is then cooled down to \SI{4}{\kelvin} in a liquid helium bath cryostat and a low temperature lithography step is performed. During this step, markers with the shape illustrated in \fig\ref{fig:markerspng}(a) are defined in the photoresist. 
After development (see microscope picture in \fig\ref{fig:markerspng}(b)) the markers are transferred on the sample surface. 
This is done, either by inductively coupled plasma reactive ion etching (ICP-RIE)(SEM image in \fig\ref{fig:markerspng}(c))
or by the deposition of metal via electron-beam evaporation (SEM image in \fig\ref{fig:markerspng}(d)). 

\subsection{Femtosecond two-photon 3D printing}

 For the fabrication of the micro-optics, a commercial 3D direct laser writing machine (Photonic Professional GT, Nanoscribe GmbH) is used. This machine is equipped with a femtosecond pulsed laser operating at wavelength of \SI{780}{\nm} and a repetition rate of \SI{80}{MHz}. A galvometric scanner is used to position the laser beam in lateral direction through a high NA microscope objective. In the vertical direction the sample is moved by a build in piezo positioning stage. During fabrication, the entire sample is first covered with liquid photoresist (IP-dip) by drop-coating. Subsequently a 63x microscope objective (NA = 1.4) is immersed into the resist (dip-in configuration). 
The aforementioned markers are then used to align 3D direct laser writing machine to the QD position. To this end, the femtosecond laser is attenuated in a way that its reflection is visible on the built-in camera, but does not expose the resist.\\

\begin{figure}[!h]
	\centering
	\includegraphics[]{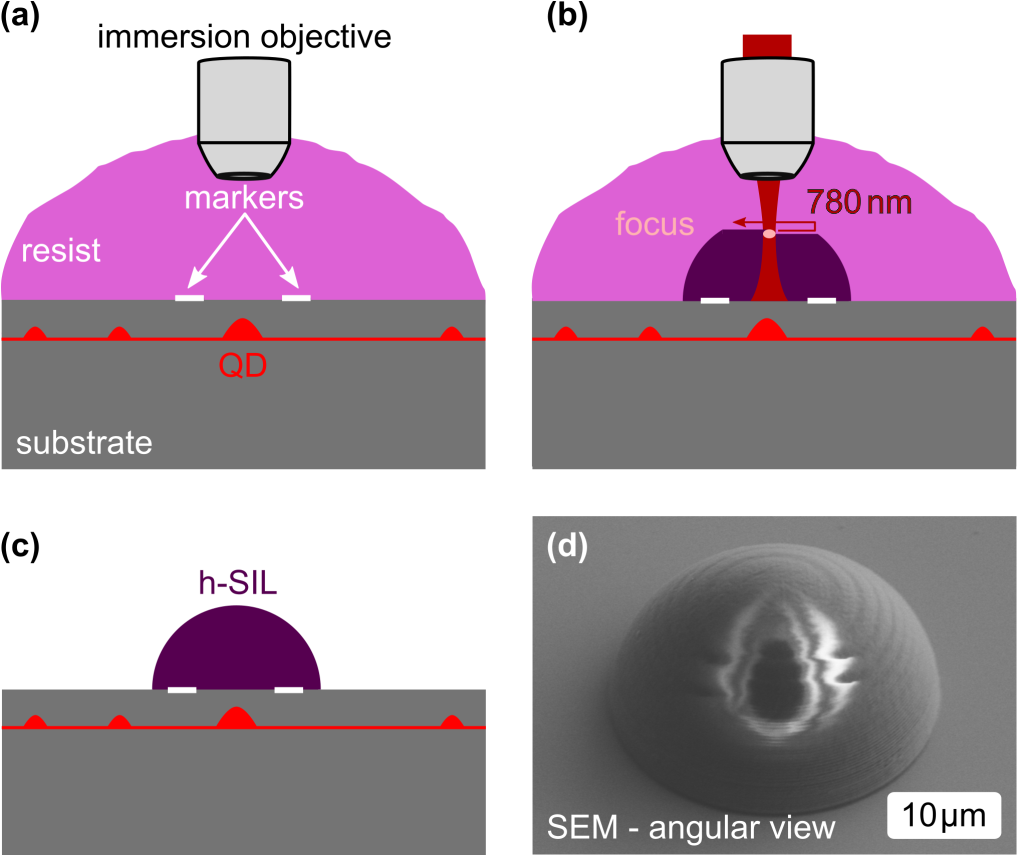}
	\caption{Schematic process flow of the 3D direct laser writing of a hemispheric SIL. \textbf{(a)} Drop coated sample with etched markers. Markers are found via inspection with the depicted immersion microscope objective. \textbf{(b)} A fs pulsed laser at \SI{780}{\nm} wavelength is focused into the resist. The resist is polymerized only in the laser focus. \textbf{(c)} Leftover resist is removed using resist developer and isopropanol. \textbf{(d)} Exemplary SEM image of the finished h-SIL. Due to the non-conductive nature of the resist, charge accumulations are  visible.}
	\label{fig:3dprinting}
\end{figure}

The coordinates (horizontal and vertical) of each marker (link of the two squares) is recorded via the piezo position readout. 
By calculating the average of all four coordinates, the writing coordinate of the structure to be exposed is determined. Ideally, this coincidences with the QD position. 
Resist exposure takes place via two-photon polymerization by the femtosecond pulsed laser \cite{fischer2013}. A CAD file serves as an input and is sliced with a spacing of \SI{100}{\nm}. 
The slices are then exposed in a layer-by-layer fashion to complete the printed structure (see \fig\ref{fig:3dprinting}(b)). 
After the exposure, leftover photoresist is removed in a developer bath (mr-Dev 600, microresist technology GmbH) for \SI{20}{\minute} and isopropanol for \SI{5}{\minute}. A sketch and an SEM picture of the completed structure (in this case a hemisphere) after development is depicted in \fig\ref{fig:3dprinting}(c) and (d). 
\fig\ref{fig:3dprinting}(d) shows charge accumulations on the lens due to the non-conducting nature of the resist.

\section*{Acknowledgements}

We acknowledge the financial support of the German Federal Ministry of Science and Education [Bundesministerium für Bildung und Forschung (BMBF)] via the projects Printoptics, Printfunction, Q.link.X 16KIS0862 and support via the project EMPIR 17FUN06 SIQUST. This project received funding from the Baden-Württemberg-Stiftung via the project Opterial. This project received funding from the EMPIR programme cofinanced by the Participating States and from the European Union’s Horizon 2020 research and innovation program. This project received funding from the European Research Council (ERC) via the projects AdG ComplexPlas and PoC 3D PrintedOptics. This project received funding from the Deutsche Forschungsgemeinschaft(DFG) via projects SPP1839 and SPP1929. This project received funding from the Center for Integrated Quantum Science and Technology (IQST).

\section*{Conflict of interests}

The authors declare no conflict of interest.

\section*{Author contribution}

M.S. conducted the deterministic fabrication and the optical measurements with the support of S.K., T.H. and S.L.P.. M.S. performed the theoretical calculations. K.W. realized the 3D printed structures and implemented the fiber mating. S.T. realized the micro-optics design with the supervision of A.H.. M.S. and T.H. performed the measurements of the fiber in-coupling with help from L.B. S.F.. S.R., A.H., P.M., S.L.P. and H.G. coordinated and supervised the project. H.G. conceived the original idea. All authors contributed to scientific discussions and preparation of the manuscript.





\medskip

%
\bibliographystyle{osajnl}
\bibliography{mybib2}

\end{document}